\documentclass[
 amsmath,amssymb,
preprint,%
]{revtex4-1}

\usepackage{graphicx}
\usepackage{dcolumn}
\usepackage{bm}

\usepackage{subcaption}

\usepackage[utf8]{inputenc}
\usepackage[T1]{fontenc}
\usepackage{mathptmx}
\usepackage{etoolbox}
\usepackage{float}
\usepackage[table]{xcolor}
\usepackage{booktabs}
\usepackage{subcaption,array}
\usepackage{diagbox}
\usepackage{multirow}%
\usepackage{slashbox}
\usepackage{geometry}
\usepackage{tikz}

\newcolumntype{a}{>{\columncolor{gray}}c}
\usepackage{natbib}

\let\oldAA\AA
\renewcommand{\AA}{\text{\normalfont\oldAA}}
\usepackage{hhline}
\makeatletter
\def\@email#1#2{%
 \endgroup
 \patchcmd{\titleblock@produce}
  {\frontmatter@RRAPformat}
  {\frontmatter@RRAPformat{\produce@RRAP{*#1\href{mailto:#2}{#2}}}\frontmatter@RRAPformat}
  {}{}
}%
\makeatother
\begin{document}

\preprint{AIP/123-QED}

\title[]{Shape transitions of RBC under oscillatory flows in microchannels}
\author{Lahcen Akerkouch}
\author{Trung Bao Le}%
 \email{trung.le@ndsu.edu.}
\affiliation{ 
Department of Civil, Construction, and Environmental Engineering\\ 
North Dakota State University\\
1410 14th N, Fargo, United States, 58102\\
}%

\date{\today}

\begin{abstract}
\textbf{ABSTRACT\\}
We investigate the dynamics of the Red Blood Cell (RBC) in microfluidic channels under oscillatory flows. The simulations employ a hybrid continuum-particle approach, in which the cell membrane and cytosol fluid are modeled using Dissipative Particle Dynamics (DPD) method, and the blood plasma is modeled as an incompressible fluid via the Immersed Boundary Method (IBM). The goal of this study is to understand the morphological modes of the RBC under transient shear rates. Our simulations show good agreement with previous experimental and computational works. Our findings demonstrate the ability to control the transient dynamics of the RBC by adjusting the oscillatory waveform at the microchannel inlet. These results suggest that oscillatory flows can be used to manipulate cells, which may have implications for cell separation and identification of pathological cells.
\end{abstract}

\maketitle

\section{Introduction}
\label{sec:introduction}
Extensive research has been conducted in the last few decades on the morphological changes of Red Blood Cells (RBCs) in fluid flows due to its importance in blood pathology \cite{Kaul1983,Barabino2010,secomb2017blood}. It has been shown that the response of RBC membrane to blood plasma dynamics can affect the overall patterns of microvascular blood flows \cite{Tomaiulo2009, Guckenberger2018, Reichel2019}. Despite a substantial body of literature, the dynamics of RBCs remain a significant challenge to be studied due to the complexity of various response modes, which result from the interaction of the suspended cellular membrane with the shear flow \cite{vlahovska2011dynamics}. 
There are several factors that can affect the dynamics of RBCs, such as the stiffness of the membrane \cite{Czaja2020}, the shear rate $\left( \bar{\dot{\gamma}}\right)$\cite{Lanotte2016,Mauer2018}, and the viscosity contrast $(\lambda)$ between the blood plasma and cytosol \citep{Mauer2018}, among other factors. As a result, RBC deformation process in shear flow is not well understood, especially under time-dependent shear rates \citep{schmidt2022oscillating, Krauss2022}. 

In free shear flows with constant shear rate $\dot{\gamma}$, the shear strength\cite{dupire2012full} is the controlling parameter of the RBC dynamics. The shape of RBCs becomes increasingly complex (more lobes) as the shear rate increases. In the range of shear rate ($\bar{\dot{\gamma}}$) from 10 $s^{-1}$ to 2,000 $s^{-1}$, the dynamics of RBCs can be classified into three main regions \cite{Lanotte2016}: (i) tumbling at weak shear rate $(\bar{\dot{\gamma}}< 10~s^{-1})$; (ii) circular/elliptical rims $(10~s^{-1} < \bar{\dot{\gamma}}< 400~s^{-1})$; and (iii) multilobes ($400~s^{-1} < \bar{\dot{\gamma}} < 2,000~s^{-1}$). In the tumbling region, the deformation was minimal and reversible, which allows the RBCs to maintain their biconcave discoid shape. As the shear rate increases to $40 s^{-1}$, the percentage of discocytes decreases and is replaced by the emergence of stomatocytes. The rolling and tumbling stomatocytes \citep{Mauer2018} appear at $\dot{\gamma} = 150~s^{-1}$ and  $250~s^{-1}$, respectively. This pattern persists up until $\bar{\dot{\gamma}} = 400~s^{-1}$ when the stomatocytes assume a shape with an elliptical rim. In the range $400~s^{-1} < \bar{\dot{\gamma}} < 2,000~s^{-1}$, RBCs with large lobes on their surface, which are referred as trilobes or hexalobes, emerge.

Studies of RBC dynamics in microchannels have shown that the RBC can transition from its biconcave discoid shape to different morphologies \citep{Noguchi2005, Fedosov2014, Guckenberger2018} under specific combinations (state diagram) of viscosity contrast, shear rate and channel confinement $(\chi)$ \citep{Mauer2018, Fedosov2014, Reichel2019}. The state diagram has revealed two main categories of RBC morphological shapes: $(i)$ symmetrical; and $(ii)$ asymmetrical types \cite{Tomaiulo2009, Quint2017, Kihm2017}. The symmetrical type contains three modes \cite{coupier2012shape}: $(a)$ bullet; $(b)$ crossaint (in rectangular channels); and $(c)$ parachute (in circular channels) shapes, while the asymmetrical type includes \cite{kaoui2009red, Reichel2019}: $(a)$ slipper; $(b)$ multilobes; $(c)$ trilobes; and $(d)$ hexalobes shapes. The shape transition in the symmetrical type has been shown to reach the stationary shape (either bullet, croissant, or parachute). However, it is still not fully clear whether or not the asymmetrical shapes are stable or they are just transient states \citep{Noguchi2005, Kaoui2011}. It has been shown that the shape transition in the symmetric type depends on the bulk flow velocity, the channel confinement, and the shear rate (the capillary number - $Ca$) \cite{kaoui2011two}. In the asymmetric mode \cite{kaoui2009red}, the shape transition mostly depends on the flow lag, which is the difference between the translation velocity of the RBC  and the velocity of plasma. In brief, it is unclear how the asymmetrical shapes emerge from the biconcave discoid shape.

Two shapes are the most frequently observed: $(i)$ the croissant shape (symmetrical) \citep{Guckenberger2018}; and $(ii)$ the slipper shape (asymmetrical). In particular, the slipper shape is characterized by the tank-treading motion of the cell membrane, which is essentially a self-rotation of the membrane around its own center of mass during the RBC propagation \citep{McWhirter2009, Guckenberger2018}. Experimental and computational studies have shown that these morphological shapes might result in distinct flow structures of blood plasma in the vicinity of the RBC \citep{Guckenberger2018}. For instance, there exists a closed vortex downstream of the RBC when the slipper shape emerges \citep{Yaya2021}. Such a vortex is absent during the croissant shape. To our knowledge, there has been no systematic effort in understanding the emergence of the extracellular flow patterns as the morphological shape of the RBC changes.

Recently, oscillatory flow (time-dependent shear rate) has been shown to be a promising technique for cell separation because cell deformation is irreversible under time-dependent shear rates \cite{mutlu2018oscillatory,schmidt2022oscillating}. Furthermore, oscillatory flows has been utilized to sort RBCs based on their size and deformability \citep{Krauss2022, schmidt2022oscillating}. Oscillatory flows can reduce the required travel distance of cells because it induces the lateral migration of cells in a short axial distance. This feature simplifies the design of microfluidic channel and thus improves the cell separation process\cite{lafzi2020inertial}. However, it is unclear on the process of morphological transition as the RBC responds to the time-dependent shear rate during this lateral migration. Therefore, it is necessary to investigate this process in details.

In this work, we utilized our hybrid continuum-particle simulation methodology \citep{akerkouch2021} to study the response of the RBC to a time-dependent shear rate. Our paper is organized as follows. First, a brief description of the numerical methods for simulating the blood plasma and the RBC is presented. Second, the obtained RBC dynamics are validated with experimental data under: $(i)$ stretching force; $(ii)$ constant shear rates (croissant and slipper shapes); $(iii)$ oscillatory shear rates. Third, we perform a parametric study where the shear rate waveform, the peak flow rate, and the initial position of the RBC were varied to induce a host of RBC morphological changes. Finally, the relationships between the RBC's shape and the extracellular flow patterns are reported as a basis for cell manipulation in future applications. 
\section{Methodology} \label{section:DPD_methodology}
\subsection{The idealized shape of the RBC}
The idealized shape of the RBC membrane is given by a set of points with coordinates $(x, y, z)$ in $3D$ space with the analytical equation \cite{Fedosov2010}:
\begin{equation}
    z = \pm D_{0}\sqrt{1 - \frac{4(x^2 + y^2)}{D_{0}^2}} \left[ a_0 + a_1\frac{x^2 + y^2}{D_{0}^2} + a_2\frac{(x^2 + y^2)^2}{D_{0}^4} \right],
    \label{eqn:RBC_Mesh} 
\end{equation}
the parameters are chosen in this work as $D_0=7.82~\mu m$ (equilibrium diameter), $a_0=0.00518$, $a_1 = 2.0026$, and~$a_2 = -4.491$. Note that the idealized shape will be used as the initial shape of the RBC membrane only. The membrane mechanics that governs the cellular deformation under loadings will be described in the following sections.

\subsection{RBC membrane model}
\label{sec:rbc_membrane_model}
As the idealized surface of the RBC membrane is known precisely according to  Equation~(\ref{eqn:RBC_Mesh}), a triangulation procedure is carried out to mimic the distribution of the spectrin links on the membrane as edges of each triangular elements (links) \citep{Fedosov2010}. A network of non-linear springs is generated for each edge to model the dynamics of the spectrin links \citep{Pivkin2008, Fedosov2010, akerkouch2021}. At each vertex $i$, the dynamics of the links are determined by the membrane force $\mathbf{F}_i^{membrane}$, which is linked to the Helmholtz's free energy $V_i$ at the same vertex $i$ through the following relationship:
\begin{equation}
    \mathbf{F}^{membrane}_{i} = - \frac{\partial V_{i}}{\partial \mathbf{r_{i}}},
    \label{eqn:RBC_membrane_forces} 
\end{equation}

with $\mathbf{r_i}$ is the position vector of the vertex $i$.

The potential $V({\lbrace \mathbf{r}_i \rbrace})$ incorporates the physical properties of the lipid bilayer: $(a)$ in-plane stretching; $(b)$ bending stiffness; $(c)$ area and volume conservation; and $(d)$ membrane viscosity
\begin{equation}
    V({\lbrace \mathbf{r}_i \rbrace}) = V_{in-plane} + V_{bending} + V_{area} + V_{volume}
\end{equation}

\subsubsection{Action potential models}
The in-plane free energy term $V_{in-plane}$ includes the elastic energy stored in the membrane modeled using the nonlinear Wormlike Chain and power $(WLC-POW)$ spring model. Here, the $WLC-POW$ potential is computed for each link $j$ formed by two vertices as,

\begin{equation}
    V_{in-plane} = \sum_{j \in 1...N_s} \left[ U_{WLC}(l_j) +  U_{POW}(l_j)\right],
    \label{eqn:V_inplane}
\end{equation}
with  $N_s$ is the total number of links forming the triangulated mesh.

The $WLC$ attractive potentials $U_{WLC}(l_j)$ for individual link $l_j$ is expressed as:
\begin{equation}
    U_{WLC} = \frac{k_{B} T l_{max}}{4p} \frac{3x^2 - 2x^3}{1 - x},
\label{eqn:WLC_Model}    
\end{equation}

where the value $x = \frac{l_j}{l_{max}}$ represents the spring deformation, in which
$l_j$, $l_{max}$, $p$, $k_{B}$, and $T$ are the length of the spring $j$, the maximum allowable length of the links, the persistence length, Boltzmann's constant, and the temperature, respectively. 

The repulsive force, described by the energy potential $U_{POW}(l_j)$, takes the form of a power function $(POW)$. The separation distance $l_j$ is a determining factor in the calculation of $U_{POW}$, which is given by:
\begin{equation}
    U_{POW}(l_j) = \frac{k_p}{(m-1)l_j^{m-1}} ~~~m>0 \text{~and~} m\ne1,
\end{equation}
where $k_p$ is the $POW$ force coefficient. The value $m = 2$ is used for the exponent \cite{Fedosov2010}.\\

The bending energy $V_{bending}$ accounting for the membrane resistance to bending is defined as,

\begin{equation}
    V_{bending} = \sum_{j \in 1...N_s} k_b\lbrack 1 - cos(\theta_j - \theta_0) \rbrack,
\end{equation}

with $k_b$, $\theta_0$ and $\theta_j$ are the bending rigidity, the spontaneous angle and the instantaneous angle between the normal vectors of two adjacent triangles sharing a common edge (link) $j$, respectively.

The area and volume conservation constraints account for the incompressibility of the lipid bilayer and the inner cytosol, respectively. They are defined as:
\begin{equation}
  \begin{aligned}
V_{area} &= \frac{k_{a}(A-A_{0}^{tot})^2}{2A_{0}^{tot}} + \sum_{k \in 1...N_t} \frac{k_{d}(A_{k}-A_{0})}{2A_0}, \\
  V_{volume} &= \frac{k_{v}(V-V_{0}^{tot})^2}{2V_{0}^{tot}},
  \end{aligned}
  \label{eqn:V_area_volume}
\end{equation}

with $N_t$, $k_a$, $k_d$, and $k_v$ are the total number of triangles, the global area, local area, and~volume constraint coefficients, respectively. $A_k$ and $A_0$ are the instantaneous area of the $k^{th}$ triangle (element) and the initial value of the average area per element. $A_{0}^{tot}$ and $V_{0}^{tot}$ are the RBC's equilibrium total area and volume, respectively. $A$ and $V$ are the instantaneous total area and total volume of the RBC. The detailed procedure to evaluate the values of $A$ and $V$ for individual elements was reported in our previous work\citep{akerkouch2021}. 

Equation (\ref{eqn:RBC_membrane_forces}) is used to calculate the precise nodal forces for each potential energy $V$ in Equations (\ref{eqn:V_inplane}) - (\ref{eqn:V_area_volume})\citep{akerkouch2021, Fedosov2010}. The internal force $\mathbf{F}^{membrane}_i$ contribution from $i^{th}$ vertex can be computed by summing all the nodal forces as:
\begin{equation}
\mathbf{F}^{membrane}_i = \mathbf{F}^{WLC}_i + \mathbf{F}^{POW}_i + \mathbf{F}^{Bending}_i + \mathbf{F}^{Area_g}_i + \mathbf{F}^{Area_{loc}}_i +  \mathbf{F}^{Volume}_i.
\label{eqn:Total_Membrane_Forces}
\end{equation}

\subsubsection{Cellular membrane/cytoskeleton interaction}
To account for the interactions between the cytoskeleton and the lipid bilayer, the bilayer-cytoskeletal interactions force $\mathbf{F}^E$  was incorporated into the total RBC membrane forces \cite{Peng2013}. In particular, $\mathbf{F}^E$ is applied when the distance between two membrane triangles with opposite normal vectors is less than the minimal activation distance $d_a = 0.2~\mu m$. The force $\mathbf{F}^E$ is applied equally to all the vertices $(i = 1, 2~\text{and}~3)$ for each of the two elements. The bilayer-cytoskeletal interactions force is given by:
\begin{equation}
        \mathbf{F}^{E}_i = k_{bs}~\mathbf{n},
    \end{equation}

 with the stiffness of the bilayer–cytoskeletal $(k_{bs} = 4.1124~pN/\mu m)$ was assumed to be in the same order of the membrane spectrin network \cite{Peng2013}. $\mathbf{n}$ is the normal vector of the triangle.

\subsection{Modeling membrane-cytosol interactions}
\label{sec:dpd_method}
The interaction between the membrane and the cytosol is modeled using the Dissipative Particles Dynamics (DPD method). DPD is a microscopic simulation technique widely used to model flow of complex fluids, in which the flow is described as group of clustered interacting particles moving as a soft lump of fluid according to the Lagrangian approach \cite{Fedosov2010}. In this work, the cytosol within the RBC is modeled using a set of randomly distributed DPD particles ($N_f$) that fill the internal volume of the cell\citep{Pivkin2008, Fedosov2010}. 

Due to the different nature of the interactions, the component of the total force of each particle $\mathbf{F_i}$ is different depending on the nature of the particle $i$ (either membrane or cytosol particle). In general, each DPD particle $i$ interact with surrounding particles $j$ within a cut-off radius $r_c$ through three pairwise additive forces: $(a)$ the conservative force $\mathbf{F}^{C}_{ij}$; $(b)$ the dissipative force $\mathbf{F}^{D}_{ij}$; and $(c)$ the random force $\mathbf{F}^{R}_{ij}$. The relative position vector between the particles $i$ and $j$ and related terms are given by: $\mathbf{r_{ij}} = \mathbf{r_{i}} - \mathbf{r_{j}}$, the distance $r_{ij} = |\mathbf{r_{ij}}|$, and the unit vector  $\mathbf{\hat{r}}_{ij} = \frac{\mathbf{r_{ij}}}{r_{ij}}$. Also, $\mathbf{v}_{i,j} = \mathbf{v}_i - \mathbf{v}_j$ is the relative velocity between the particles $i$ and $j$ with velocities $\mathbf{v}_i$ and $\mathbf{v}_j$.

For a DPD particle $i$ of the cytosol fluid, the total force $\mathbf{F_i}$ is: 
\begin{equation}
   \mathbf{F_i} = \sum_{j\ne i} \mathbf{F}^{C}_{ij} + \mathbf{F}^{D}_{ij} + \mathbf{F}^{R}_{ij}.
 \end{equation}

For the membrane particles, the total force $\mathbf{F}_i$ acting on each membrane particle is given by the sum of the membrane force $\mathbf{F}^{membrane}_i$, the bilayer-cytoskeletal interactions force $\mathbf{F}^E$ and the contributing forces from the surrounding DPD fluid particles from the cytosol:
\begin{equation}
   \mathbf{F_i} = \mathbf{F}^{membrane}_i + \mathbf{F}^E_i + \sum_{j\ne i} \mathbf{F}^{C}_{ij} + \mathbf{F}^{D}_{ij} + \mathbf{F}^{R}_{ij}.
 \end{equation}

 The mathematical formulation of the conservative force $\mathbf{F}^{C}_{ij}$, the dissipative force $\mathbf{F}^{D}_{ij}$, and the random force $\mathbf{F}^{R}_{ij}$ for the membrane and the cytosol fluid particles are explained below.

\subsubsection{The conservative force}
The conservative force $\mathbf{F}^{C}_{ij}$ is given by :

\begin{equation}
      \begin{aligned}
\mathbf{F}^{C}_{ij} &= F^{C}_{ij}(r_{ij}) \mathbf{\hat{r}}_{ij},\\
\mathbf{F}^{C}(r_{ij}) &=  
     \begin{cases}
       a_{ij} \left( 1 - \frac{r_{ij}}{r_c}\right) &\quad\text{for}~~r_{ij}\le ~r_c,\\
       0 &\quad\text{for}~~r_{ij} > ~r_c,\\
     \end{cases}
      \end{aligned}
\end{equation}

where $a_{ij} = 20$ is the conservative force coefficient between particles $i$ and $j$. Note that the particle $i$ and $j$ can be either membrane or cytosol fluid particle. Thus, there are two types of interactions: i) cytosol fluid/fluid; and ii) membrane/fluid particle interactions \cite{Fedosov2010}.
 
\subsubsection{The dissipative force}
The dissipative force $\mathbf{F}^{D}_{ij}$ for the membrane particles is computed as:

\begin{equation}
    \mathbf{F}_{ij}^{D} = -\Gamma^T \mathbf{v}_{ij} - \Gamma^C \left(\mathbf{v}_{ij} \cdot \mathbf{\hat{r}}_{ij}\right) \mathbf{\hat{r}}_{ij}.
    \label{eqn:Membrane_Dissipative_Viscosity}
\end{equation}

The membrane viscosity is a function of both dissipative parameters, $\Gamma^T$ and $\Gamma^C$. The superscripts $T$ and $C$ denote the translational and central components. Here, $\Gamma^T$ is responsible for a large portion of the membrane viscosity in comparison to $\Gamma^C$. In addition, $\Gamma^C$ is assumed to be equal to one third of $\Gamma^T$ in Equations \eqref{eqn:Membrane_Dissipative_Viscosity}\citep{Fedosov2010}. Consequently, these parameters relate to the physical viscosity of the membrane $\eta_m$ as:
\begin{equation}
\left\{\begin{aligned}  \eta_m = \sqrt{3}\Gamma^T &+ \frac{\sqrt{3}\Gamma^C}{4}, \\ 
\Gamma^C = \frac{\Gamma^T}{3}.
\end{aligned}\right.
\label{eqn:gamma_assumption}
\end{equation}

Hence, the dissipative force $\mathbf{F}_{ij}^{D}$ of the membrane particles can be expressed as:
\begin{equation}
\mathbf{F}_{ij}^{D} = - \frac{12}{13\sqrt{3}}\eta_m \mathbf{v}_{ij} - \frac{4}{13\sqrt{3}}\eta_m \left(\mathbf{v}_{ij}\cdot \mathbf{\hat{r}}_{ij}\right) \mathbf{\hat{r}}_{ij}.
\label{eqn:Membrane_Dissipative_Viscosity_Simplified}
\end{equation}

The dissipative force $\mathbf{F}^{D}_{ij}$ for the cytosol fluid particles is defined as:
\begin{equation}
\mathbf{F}^{D}_{ij} = -\gamma \omega^D (r_{ij}) ( \mathbf{v}_{ij} \cdot \mathbf{\hat{r}}_{ij}) \mathbf{\hat{r}}_{ij},\\
\end{equation}

the quantity $\gamma$ is a constant coefficient defining the strength of the dissipative force. The weight functions, $\omega^D(r_{ij})$ and $\omega^R(r_{ij})$ are given by:
\begin{equation}
    \omega^D(r_{ij}) = \left[ \omega^R(r_{ij}) \right]^2, 
\end{equation}

\begin{equation}
\omega^{R}(r_{ij}) =  
     \begin{cases}
       \left( 1 - \frac{r_{ij}}{r_c}\right)^s &\quad\text{for}~~r_{ij}\le ~r_c,\\
       0 &\quad\text{for}~~r_{ij} > ~r_c,\\
     \end{cases}
\end{equation}

with $s = 1$ following the original DPD method \citep{Fedosov2010}. However, other works revealed that the decrease of this parameter $s = 0.5$ to $0.25$ increases the viscosity of the DPD fluid \citep{Fan2006}. The particle $i$ represent the fluid particle, while the particle $j$ can be fluid or membrane particle within the cut-off radius $r_c$.

\subsubsection{The random force}
Using the assumptions in Equation (\ref{eqn:gamma_assumption}), the random force for membrane particles can be simplified as:
\begin{equation}
    \mathbf{F}_{ij}^{R} = \sqrt{2 k_B T} \left(2\sqrt{\frac{2\sqrt{3}}{13}\eta_m}~d\overline{W^{S}_{ij}}\right)\mathbf{\hat{r}}_{ij},
\label{eqn:Membrane_Random_Viscosity_Simplified}
\end{equation}

where $tr(d\mathbf{W}_{ij})$ is the trace of the random matrix of independent Wiener increments $d\mathbf{W}_{ij}$, and $d\overline{\mathbf{W}^{S}_{ij}} = d\mathbf{W}^{S}_{ij} - \frac{tr(d\mathbf{W}^{S}_{ij})}{3}$ is the traceless symmetric part. 

The random force $\mathbf{F}^{R}_{ij}$ for the cytosol fluid are defined as:
\begin{equation}
\mathbf{F}^{R}_{ij} = \sigma \omega^R (r_{ij}) \cdot \frac{\vartheta_{ij}}{\sqrt{dt}} \cdot \mathbf{\hat{r}}_{ij}, ~~~\sigma^2 = 2\gamma k_{B}T,
\end{equation}
where $\sigma$ is a constant coefficient defining the strength of the random force, $dt$ is the physical time step , $\vartheta$ is a normally distributed random variable with zero mean and unit variance and $\vartheta_{ij}=\vartheta_{ji}$. Note that the particle $i$ and $j$ must be both cytosol fluid particles.

\subsubsection{Plasma and cytosol viscosity contrast} \label{sec:dpd_to_physical}
At the physiological blood conditions, the viscosity ratio between the blood plasma and the RBC cytosol is equal to $5.0$ ($\lambda = \frac{\mu_{cytosol}}{\mu_{plasma}} = 5.0$) \citep{Wells1969}. To ensure that this condition is met, the dynamic viscosity of the plasma is set to be $\mu_{plasma}  = 1.2~mPa~s$. The viscosity condition is enforced on the cytosol fluid by calibrating the parameters of the dissipative and the random forces \citep{Duy2020} (e.g  $\gamma$ and $\sigma$). Specifically, the dynamic properties of the DPD particles of the cytosol fluid are given in the dimensionless DPD unit as \cite{Fan2006}:
\begin{equation}
    \begin{aligned}
       \textrm{mass diffusivity:~~~~}  D_f = \frac{45 k_B T}{2 \pi \gamma \rho r_c^3},  \\
        \textrm{dynamics viscosity:~~~~}  \mu = \frac{\rho D_f}{2} + \frac{2 \pi \gamma \rho^2 r_c^5}{1575},
    \end{aligned}
\end{equation}

with $\rho$ is the density. The DPD dimensionless parameters and physical units are linked \cite{Ghoufi2013} in order to compute the coefficients of the dissipative and randoms forces for the cytosol dynamic viscosity of $\mu_{cytosol} = 6~mPa~s$, based on the viscosity ratio $\lambda = 5$. The details of the conversion procedure are summarized in Table \ref{tab:dpd_fluid_visc}.

\subsection{Scaling of model and physical units} \label{sec:Scaling}
One challenge in DPD modeling is establishing a relationship between the modeled quantities and the physical values \citep{Ye2019}. Since this relationship is not explicit, it is necessary to use a scaling argument to recover this relationship \citep{Fedosov2010, peng2013lipid}. For each parameter, the superscript $M$ and $P$ corresponds to the model and physical units, respectively. 

The length scale $r^M$ is defined as:
\begin{equation}
    r^M = \frac{D_0^{P}}{D_0^{M}}~(m),
\end{equation}

The energy per unit mass $k_{B}T$ and the force $N$ scaling values are given by:
\begin{equation}
  \begin{aligned}
(k_{B}T)^M &= \frac{Y^P}{Y^M}\left(\frac{D_0^{P}}{D_0^{M}}\right)^2(k_{B}T)^P, \\
  N^M &= \frac{Y^P}{Y^M}\frac{D_0^{P}}{D_0^{M}}N^P,
  \end{aligned}
\end{equation}
with $Y$ is the membrane Young’s modulus. 

The timescale $\tau$ is defined as following:
\begin{equation}
    \tau = \left( \frac{D_0^{P}}{D_0^{M}} \frac{\eta_m^P}{\eta_m^M}\frac{Y^M}{Y^P} \right)^\alpha,
\label{eqn:Time_Scale}    
\end{equation}

with $\alpha = 1$ is the scaling exponent.
\subsection{Coarse-graining Procedure}
\label{sec:coarse_graining}
A full-scale model of a RBC typically consists of millions of particles, which are required to accurately simulate protein dynamics \citep{tang2017openrbc}. However, it is not feasible to use such a full-scale model in a Fluid-Structure Interaction (FSI) simulation due to the high computational cost.  We followed the coarse-graining procedure of Pivkin et al. (2008) \citep{Pivkin2008} to represent the RBC membrane by a smaller number of particles (coarse-grained model). This procedure does not allow a detailed simulation of separate proteins, but it is versatile enough to capture the overall dynamics of the RBC membrane. The parameters of the coarse-grained model ($c$) are computed from the ones of the fine-scaled model $(f)$ by a scaling procedure. The examples of such paramemters are explained below.

Based on the equilibrium condition, Pivkin et al.~\cite{Pivkin2008} proposed a coarse-graining procedure based on the area/volume constraint for the spring equilibrium $l_0$ and maximum $l_{max}$ lengths as follows:

\begin{equation}\label{eq:l_0_l_m_CG}
 l_0^c = l_0^f \sqrt{\frac{N_{v}^f - 2}{N_{v}^c - 2}}
   \quad\text{and}\quad 
l_{max}^c = l_{max}^f \sqrt{\frac{N_{v}^f - 2}{N_{v}^c - 2}},
\end{equation}

the role of $l_0$ and $l_{max}$ is critical in determining the response from the WLC model as seen in Equation~(\ref{eqn:WLC_Model}), with $l_{max} = 2.2~l_0$ in the fine-scaled model. Due to the scaling in Equation~(\ref{eq:l_0_l_m_CG}), the~value of $x_0 = \frac{l_0}{l_{max}} = \frac{1}{2.2}$ does not change as the model is coarse-grained from the number of vertices $N_v^f$ to $N_v^c$. 

Furthermore, as the number of vertices reduces, the average angle between the pairs of adjacent triangles increases. Therefore, the spontaneous angle $\theta$ is adjusted accordingly in the coarse-grained model as:
\begin{equation}\label{eq:theta_CG}
  \theta_0^c = \theta_0^f \frac{N_{v}^f}{N_{v}^c}
   \quad\text{with}\quad 
\theta_0^f = \arccos \left( \frac{\sqrt{3}(N_v^f -2) - 5\pi}{\sqrt{3}(N_v^f -2) - 3\pi} \right).
 \end{equation}
 
To~maintain the shear and area-compression moduli, the~parameters $p$ and $k_p$ are adjusted as:
\begin{equation}
 p^c = p^f \frac{l_{0}^f}{l_{0}^c}
   \quad\text{and}\quad 
k_p^c = k_p^f \left(\frac{l_{0}^c}{l_{0}^f}\right)^{m+1}.
\label{eq:p_k_p_CG}
\end{equation}

\subsection{Time integration}
In this work, we implemented the modified Velocity-Verlet algorithm \cite{Groot1997}, which consists of two primary steps. The first step involves determining the new position of the particle $i$ ($\mathbf{r}_i$) while predicting the velocity ($\tilde{\mathbf{v}_i}$), and the second step involves correcting the velocity by utilizing the computed force ($\mathbf{F}_i$) based on the predicted velocity and the new position as follows.
 
\begin{align}
  \mathbf{r}_i(t + dt) &= \mathbf{r}_i(t) +  dt \mathbf{v}_i(t) + \frac{1}{2} dt^2 \mathbf{F}_i(t), \nonumber \\
  \tilde{\mathbf{v}_i}(t + dt) &= \mathbf{v}_i(t) + \Lambda dt \mathbf{F}_i(t),\\
  \mathbf{F}_i(t + dt) &= \mathbf{F}_i(\mathbf{r}_i(t + dt), \tilde{\mathbf{v}_i}(t+dt)), \nonumber\\
  \mathbf{v}_i(t + dt) &= \mathbf{v}_i(t) + \frac{1}{2} dt(\mathbf{F}_i(t) + \mathbf{F}_i(t + dt)), \nonumber
\end{align}
where $\tilde{\mathbf{v}_i}(t + dt)$ is the predictive velocity at time $t + dt$ and $\Lambda$ is the variable which accounts for the effects of the stochastic processes. The value of $\Lambda$ is chosen to be the optimal value  \cite{Groot1997} $\Lambda = 0.65$.

\subsection{Fluid-Structure Interaction simulation of RBC in flows}
\subsubsection{Numerical methods}
The blood plasma was considered as an incompressible Newtonian fluid modeled using the incompressible three-dimensional unsteady Navier-Stockes equations, with density $\rho $ and kinematic viscosity $\nu = \frac{\mu_{plasma}}{\rho}$. The governing equations (continuity and momentum) read in Cartesian tensor notation as follows ($i=1,2,3$ and repeated indices imply
summation):
\begin{eqnarray}
\frac{\partial {u_{i}}}{\partial {x_{i}}} &=&0,  \label{eqn:NS} \\
\frac{\partial {u_{i}}}{\partial {t}}+\frac{\partial {(u_{i}u_{j})}}{%
\partial {x_{j}}} &=&-\frac{\partial {p}}{\partial {x_{i}}}+\nu \frac{%
\partial ^{2}{u_{i}}}{\partial {x_{j}}\partial {x_{j}}}.
\end{eqnarray}%
In the above equations, $u_{i}$ is the $i^{th}$ component of the velocity
vector $\mathbf{u}$; $t$ is time; $x_{i}$ is the $i^{th}$ spatial coordinate; $p$ is the
pressure divided by $\rho $. The characteristic velocity scale is chosen as $U_{0}$. The length scale $L_s$ is set to equal $8~\mu m$ for all cases. Note that this length scale is chosen to reflect the diameter of the RBC at the equilibrium condition.

The fluid solver is based on the sharp-interface curvilinear-immersed boundary (CURVIB) method in a background curvilinear domain that contains the RBC model \citep{Ge2007}. The CURVIB method used here has been applied and validated in various FSI problems across different biological engineering areas \citep{Le2010, Le2012, Le2019}. In our previous work \citep{akerkouch2021}, we utilized the capabilities of the CURVIB method to capture accurately the complex cellular dynamics of the RBC in fluid flows. 

The dynamics of the RBC in flow is thus simulated with a hybrid continuum-particle approach since the Fluid-Structure Interaction (FSI) methodology involves the coupling of DPD methods and the solvers for Navier-Stokes equations. The details of the FSI procedures are reported in our previous works \cite{Ge2007, akerkouch2021}.

\subsection{Computational setups} \label{sec:FSI_configs}
\begin{figure}
    \centering
    \includegraphics[width=1\linewidth]{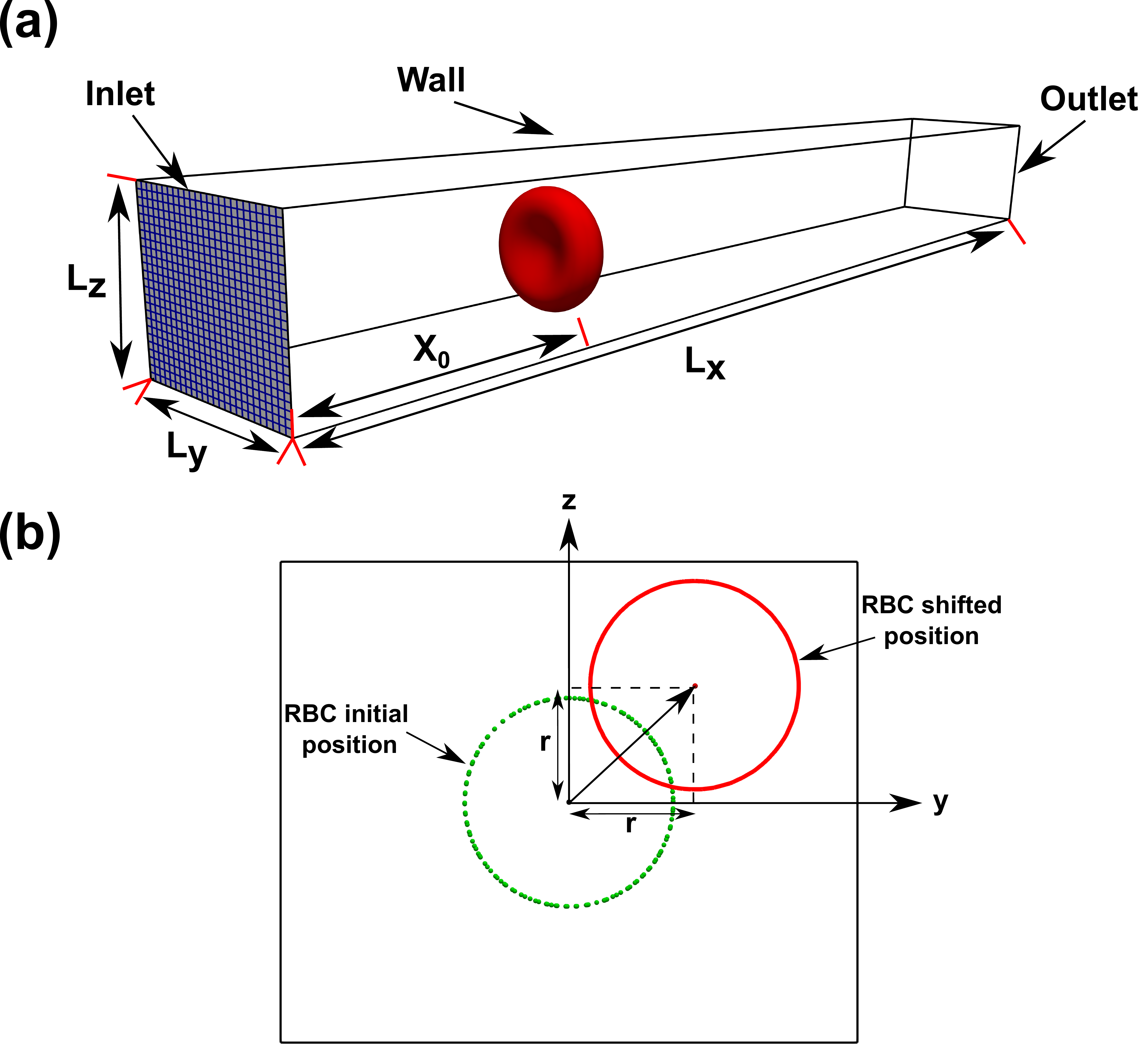}
    \caption{$(a)$ The computational setup for the FSI simulation of a single RBC in a rectangular channel of size $L_x \times L_y \times L_z$. The inlet plane is shown in blue, which shows uniform grid lines to illustrate the computational mesh. The RRBC is placed at an axial distance $x_0$ from the inlet plane. $(b)$ The sketch of the cross-section of the computational domain to illustrate the definition of the radial shift step $(r)$. The dash line shows that the RBC is placed at the channel's center-line. The solid line depicts how the cell is tranversely shifted from the cross-sectional center along the bisector the first quadrant by a radial shift ($r$) in the $y - z$ plane (Table \ref{tab:summary_factors}). }
    \label{fig:simulation_setup}
\end{figure}

Fluid-Structure Interaction simulations are performed to determine the dynamics of RBC in a confined micro-channel \citep{Guckenberger2018}. The computation domain is defined as a rectangular channel containing a single RBC as illustrated in Figure \ref{fig:simulation_setup}a. The dimensions of the domain along the $x,y$, and $z$ are $L_x$ (the length), $L_y$ (the width) and $L_z$ (the height), respectively. The computational domain is discretized as a structured grid of size $N_i \times N_j \times N_k$ with the spatial resolution in three directions $(i,j,k)$ are $ \Delta x \times \Delta y \times \Delta z$, respectively. The details of the channels used in the simulations are listed in the Table \ref{tab:Channels}.

The RBC locates initially at $t = 0$ in a axial distance of $x_0$ from the inlet. The transverse location of the RBC is placed along the bisector of the first quadrant with a radial shift ($r$). Thus, the tranverse coordinates of the RBC are $y_0 = r$ and $z_0 = r$, respectively as shown in Figure \ref{fig:simulation_setup}b ($r$ is the radial shift). With this configuration, the RBC confinement is defined as  the ratio between the effective RBC diameter $D_r = \sqrt{\frac{A_0^{tot}}{\pi}}$ and the domain height $L_z$:
\begin{equation}
    \chi = \frac{D_r}{L_z}.
\end{equation}

The initial shape of the RBC is first set to be the idealized shape (Equation \ref{eqn:RBC_Mesh}) for all simulation cases at the initial time $t = 0$. A short period of relaxation $t_{relax}$ is allowed for the RBC under no external load (no flows) so that the internal forces of the RBC membrane balance. An uniform flow velocity $U(t)$ is then applied at the channel inlet at $t > t_{relax}$ to induce the RBC's deformation depending on the controlling strategy.  The average shear rate across the channel height is defined as the ratio between the bulk velocity $U(t)$ and the domain's height:
\begin{equation}
 \overline{\dot{\gamma (t)}} = \frac{U(t)}{L_z}
 \label{eqn:shear_rate}
\end{equation}

\subsubsection{Constant shear rate condition ($I_0$)}
\label{sec:constant_shear_rate}
Following  the experimental study of Guckenberger et al. \cite{Guckenberger2018} (Channel-1, Table \ref{tab:Channels}), FSI simulations of a RBC in channel flow with a constant flow rate are carried out with $x_0 = 22.5~\mu m$. To highlight the constant flow rate, the notation $I_0$ is introduced to emphasize this condition. As shown in Table \ref{tab:validation_summary}, a constant inflow velocity $U(t) = \psi_0$ are required at the inlet of the computational domain. Two values of $\psi_0$ are considered: $(i)$ $\psi_0 = U_3 = 2~mm/s$; and $(ii)$ $\psi_0 = U_4 =  6~mm/s$. In these cases, two values of the radial shift are also investigated: $r_1 = 0$ and $r_3 = 0.7~\mu m$. To simplify the discussions, the numerical values for the bulk velocity $\psi_0$ will not be explicitly referred to. Instead, only the acronyms ($U_3$ and $U_4$) will be used for reasons that will be evident in the subsequent texts. 

Using these notations, the FSI simulation cases are named using the convention for each type of inflow waveform ($I$), the bulk velocity ($U$), the radial shift ($r$), and the channel type, respectively. The first case ($I_0U_3r_1 \chi_1$) is configured with $\left(\psi_0 = U_3 = 2~mm/s\right)$ and $r = r_1 = 0~\mu m$. The second case ($I_0U_4r_3 \chi_1$) is carried out with $\psi_0 = U_4 = 6~mm/s$ and $r = r_3 = 0.7~\mu m$. Here, the Reynolds number is defined as $R_e = \frac{U L_s}{\nu}$. The kinematic fluid viscosity of blood plasma is chosen as $\nu = \frac{\mu_{plasma}}{\rho}= 1.2 \times 10^{-6}~m^2 /s$. The summary of the parameters for each simulation case is shown in Table \ref{tab:validation_summary}.

First, $t_{relax} = 10~ms$ and $7.0~ ms$ are set for $I_0U_3r_1 \chi_1$ (croissant) and $I_0U_4r_3 \chi_1$ (slipper) simulations, respectively. After the relaxation period, a linear ramping period is set for each simulation case $t_{ramp} = 30~ ms$ and $20~ms$ are set for $I_0U_3r_1 \chi_1$ and $I_0U_4r_3 \chi_1$. During this ramping period, the bulk velocity $U(t)$ is linearly increased. The value of $U(t)$ reaches $\psi_0$ at the end of the ramping period.

\subsubsection{Stepwise oscillatory flows ($I_s$)}
\label{sec:oscillatory_shear_rate}
\begin{figure}
\vspace{-3cm}
    \centering  \includegraphics[width=0.8\linewidth]{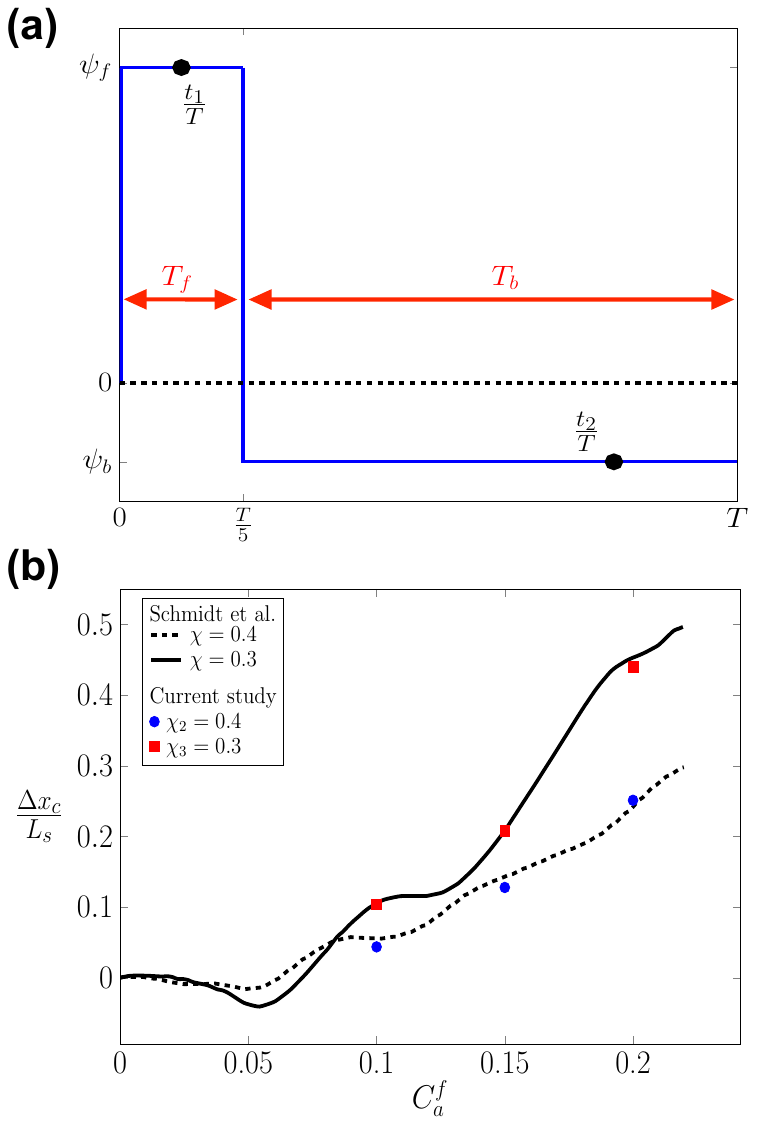}
    \caption{ The propulsion step $\Delta x_c$ (Equation \ref{eqn:propulsion_step}) as a function of the forward capillary number $Ca^f$ (Equation \ref{eqn:capillary_number}). The propulsion step $\Delta x_c$ is shown in term of the length scale $(L_s =8 \mu m)$. $(a)$ The bulk flow waveform of the inflow ($U(t)$) has a stepwise shape (see Equation \ref{eqn:stepwise_waveform}). Two time instances ($\frac{t_1}{T}$ and $\frac{t_2}{T}$) are shown to exemplify the changes of RBC shapes over time. $(b)$ Three values of $Ca^f = 0.1, 0.15$, and $0.2$ (red squares and blue circles) are simulated. The computed values of $\Delta x_c$ are compared with the previous results of Schmidt et al. (2022) \cite{schmidt2022oscillating} (solid lines). }
    \label{fig:cfd_validation}
\end{figure}

To further validate our FSI model in oscillatory flows, the propulsion of the RBC in square channels is investigated \cite{schmidt2022oscillating}. Two square channels (Channel-2 and Channel-3) with side lengths $L_z = 16~\mu m~\text{and}~21~\mu m$ are used for the simulations, resulting in confinements $\chi_2 = 0.4$ and $\chi_3 = 0.3$, respectively. The initial location of the RBC is on the channel axis ($x_0 = 16~\mu m$, $r = r_1 = 0$). The computational configuration including the grid spacing, RBC surface meshes, and boundary conditions are shown in the Figure \ref{fig:simulation_setup}a and Table \ref{tab:Channels}. 
A stepwise asymmetric oscillatory waveform $I_s$ is used with two phases: $(i)$ forward $T_f$; and $(ii)$ backward $T_b$  periods $\left(\frac{T_b}{T_f} = 4\right)$ as shown in Figure \ref{fig:cfd_validation}a. The velocities during the forward and backward phases are $\psi_f$ and $\psi_b$  $\left(- \frac{\psi_f}{\psi_b} = 4\right)$, respectively. The formula for the waveform is defined as:
\begin{equation}
U(t) =  
     \begin{cases}       
       \psi_f &\quad\text{for} ~~ 0 \leq t \le ~ \frac{T}{5},\\
       \psi_b &\quad\text{for}~~ \frac{T}{5} \leq t \leq T \\
     \end{cases}
     \label{eqn:stepwise_waveform}
\end{equation}
Following this formula, the flow has a forward phase ($\psi_f > 0$) and a backward flow phase ($\psi_b < 0$). The maximum shear rate is defined as ($\overline{\dot{\gamma_f}} = \frac{\psi_f}{L_z}$).

The Capillary number $Ca^f$ in the forward flow phase is given as:
\begin{equation}
    Ca^f = \frac{4 \psi_f L_s t_R}{L_z^2},
    \label{eqn:capillary_number}
\end{equation} with $t_R = \frac{L_s \mu_{plasma}}{2 \mu_0}$.

There are 6 values of $\psi_f$ are examined as $\psi_1 = 1.05$, $\psi_2 = 1.56$, $\psi_3 = 2.1$, $\psi_4 = 2.17$, $\psi_5 =  3.25$, and $\psi_6 = 4.35$ mm/s, respectively. Following the naming convention of the simulations, six cases are formed with the respective parameters: $I_s \psi_1 r_1 \chi_2$; $I_s \psi_2 r_1 \chi_2$; $I_s \psi_3 r_1 \chi_2$; $I_s \psi_4 r_1 \chi_3$, $I_s \psi_5 r_1 \chi_3$, $I_s \psi_6 r_1 \chi_3$ as shown in Table \ref{tab:validation_summary}. As the waveform applied was of a stepwise nature, with a gradual increase, there was no relaxation time taken into consideration for this case $(t_{relax} = 0)$.

Under these oscillatory conditions, the axial propulsion step $(\Delta x_c)$ is recorded at the end of the forward time interval of the asymmetric oscillating flow ($t = T_f = \frac{T}{5}$), as a function of the forward (peak) capillary number $Ca^f$ for the chosen shear rates  \cite{schmidt2022oscillating}. Thus $\Delta x_c$ is defined as the displacement of the RBC's centroid ($C$) at the end of the forward phase ($t = \frac{T}{5}$):
\begin{equation}
    \Delta x_c = x_c(t=\frac{T}{5}) - x_c(t=0)
    \label{eqn:propulsion_step}
\end{equation}

\subsubsection{Sinusoidal flow simulations}
\label{sec:sinusoidal_flow_simulations}
\begin{figure}[H]
\vspace{-3cm}
    \centering
\includegraphics[width=0.6\linewidth]{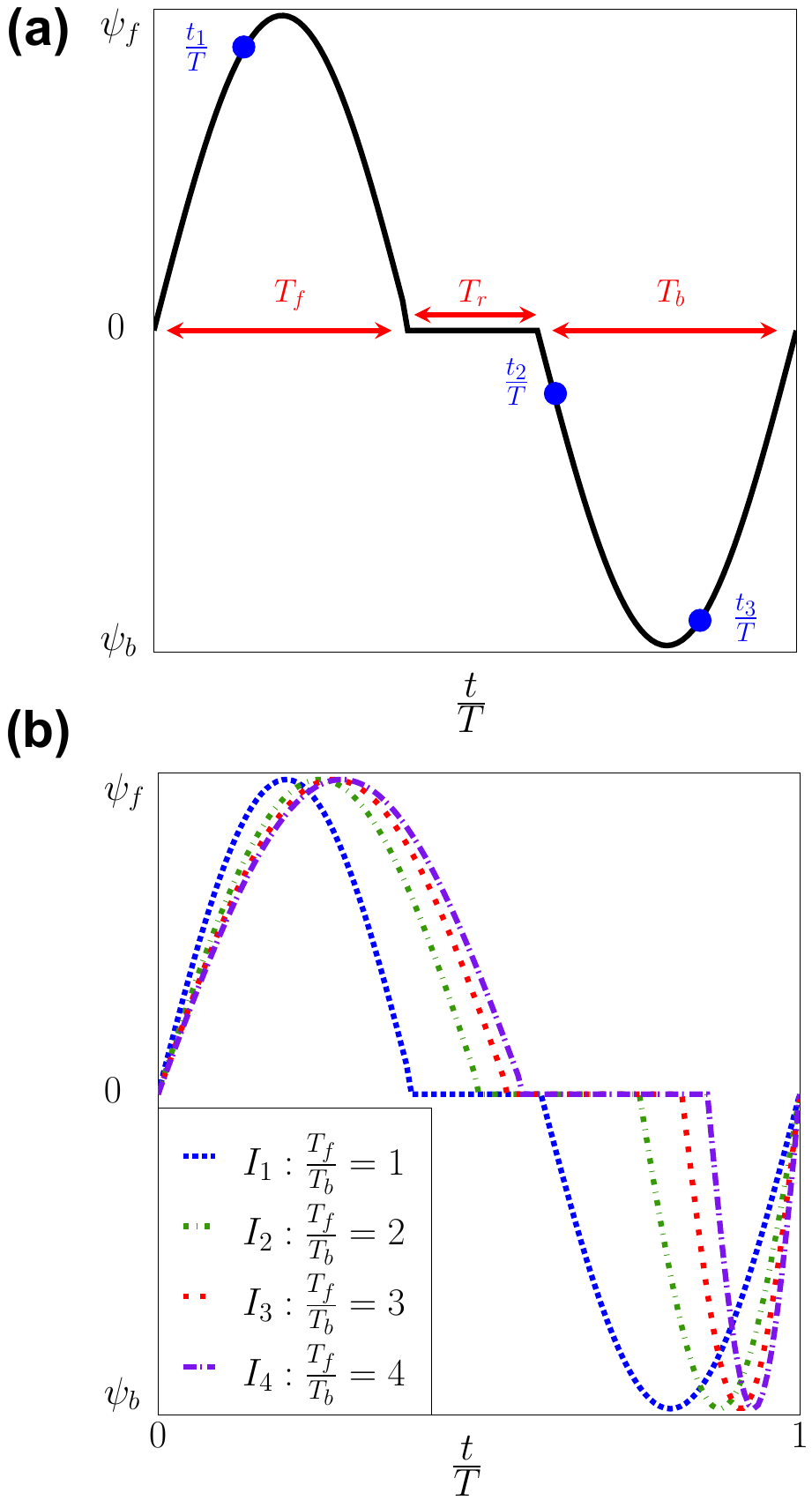}
    \caption{$(a)$ Oscillating time-dependent inflow velocity $U(t)$ profile with forward $(T_f)$, rest $(T_r)$ and backward $(T_b)$ time intervals. $(b)$ Four inflow types with different $\frac{T_f}{T_b}$ rations were considered to test flows with longer forward phase and shorter backward phase. $T_r$ is defined as half the forward time period and was introduced to minimize the numerical instabilities when the RBC is transitioning from the forward to the backward motion. The time instances $\frac{t_1}{T}$, $\frac{t_2}{T}$ and $\frac{t_3}{T}$ shown in $(a)$ represent an example of the time sampling at which the RBC shapes were recorded for each waveform. The exact values of the time instances are shown in Figure \ref{fig:RBC_inflow_shapes} and Tables \ref{tab:I1}-\ref{tab:I4}.}
    \label{fig:waveform_types}
\end{figure}

To study the effect of the pulsatile flow on the propulsion and the behavior of the cellular response (morphology changes) of the RBC, we considered time-periodic flow $U(t)$. The flow time period consists of three separate phases: $(i)$ the forward $(T_f)$; $(ii)$ the resting $(T_r)$, and the backward $(T_b)$ periods, with $T = T_f + T_r + T_b = 50~ms$ $(f = 20~Hz)$. The asymmetry of the waveform is adjusted by changing the values of $T_f$, $T_r$, and $T_b$. 
The formula for the waveform is:
\begin{equation}
U(t) =  
     \begin{cases}       
       A \sin(2 \pi \frac{t}{T_f}) &\quad\text{for} ~~ 0 \leq t \le ~T_f,\\
         0                        &\quad\text{for}~~ T_f \leq t \leq T_f + T_r \\
       A \sin(2 \pi \frac{(t - T_f - T_r)}{T_b}) &\quad\text{for}~~ T_f + T_r \leq t \leq T \\
     \end{cases}
\end{equation}
The reversible waveform ($I_1$) is created with $T_f = T_b$ (completely symmetric). The irreversible waveforms ($I_2$, $I_3$, and $I_4$) are formed by progressively reducing the period of $T_b$. Four distinct inflow types were generated with symmetry and asymmetric waveforms ($I_1$, $I_2$, $I_3$, and $I_4$) as seen in Figure \ref{fig:waveform_types} and Tables \ref{tab:physical_time} and \ref{tab:summary_factors}. For each of these waveforms, three different velocity magnitudes ($A$ = $U_1$, $U_2$ and $U_3$) were considered. Furthermore, three different radial shift ($r_1$, $r_2$ and $r_3$) were chosen for simulations. In total, the combinatoric arrangements lead to a total of 36 distinct simulation cases with the notation $I_m U_n r_p \chi_1$ with the corresponding values of the indices $m = 1, 2, 3, 4$, $n= 1 ,2 ,3$, and $p= 1, 2, 3$. The outline of the simulation cases are shown in Table \ref{tab:all_our_oscillaotry_cases}. In addition, the RBC shapes are recorded over a time period of two cycles $2T$ as exemplified in Figure \ref{fig:waveform_types}b, in which the initial location of the RBC is set at $x_0 = 22.5~\mu m$. Due to the nature of the sinusoidal waveform applied, there was no relaxation time for all of these cases $(t_{relax} = 0)$. In this case, the centroid's displacement is monitored continuously as the function of time:
\begin{equation}
    \Delta x_c (t) = x_c(t) - x_c(t=0)    
\end{equation}

\section{Results}
\subsection{Model validation}
\subsubsection{Coarse-graining validation}
\begin{figure}[H]
    \centering
    \vspace{-2cm}
    \includegraphics[width=0.75\linewidth]{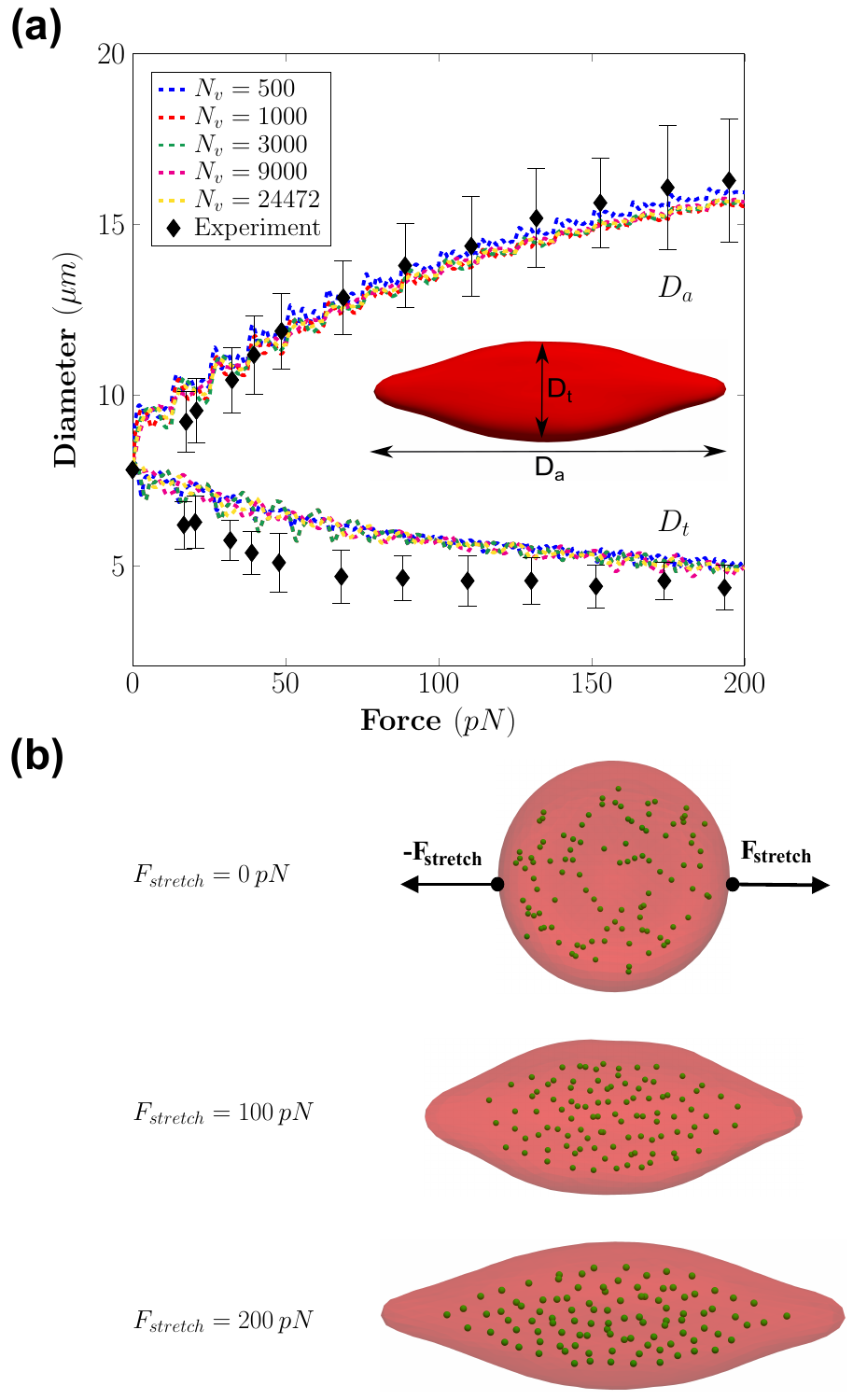}    
    \caption{$(a)$ The recorded axial $(D_a)$ and transverse $(D_t)$ diameters of the RBC response under incremental stretching force for different coarse-graining levels compared with experiment \citep{Mills2004}. $(b)$ The deformation response of the RBC membrane and the cytoplasm (green particles) under the stretching force $\mathbf{F_{stretch}}$.}
    \label{fig:streching_test}
\end{figure}

To first validate the coarse-graining procedure employed in our study, a stretching test is carried out and aimed to replicate the experimental test of Mills et al. (2004)  \citep{Mills2004}. In this experiment, two external forces $\mathbf{F}_{stretch}$ with an opposite direction are applied on both sides of the RBC. The magnitude of the force $\mathbf{F}_{stretch}$ is increased in a stepwise manner from $0$ to $200~pN$ (a total of 16 steps). The axial diameter $(D_a)$ and transverse diameter $(D_t)$ were measured for every step. $D_a$ refers to the diameter in the direction of stretch, while $D_t$ is the diameter measured in the direction orthogonal to the stretch. The definitions of $D_a$ and $D_t$ are shown in Figure \ref{fig:streching_test}a. The simulations were performed systematically with different RBC surface mesh resolutions by changing the number of vertices ($N_v$).  The parameters to describe the physical characteristics of the RBC are listed in Table \ref{tab:RBC_parameters}. Following the coarse-graining procedure, the model parameters parameters for the cell membrane such as the equilibrium length, the persistence length, the spring stiffness, and the spontaneous angle are computed for each value of $N_v$ as in Table \ref{tab:RBC_DPD_Parameters}. The cytosol fluid is modeled by a set of particles $N_f = 100$, which locate within the interior volume of the cell membrane as shown in Figure \ref{fig:streching_test}b. 

The current RBC model accurately replicates the elastic response of the RBC under stretching forces, as revealed by the results shown in Figures \ref{fig:streching_test}.  During membrane stretching under the external stretching force from $0$ to $200~pN$, the dynamic response of cytosol particles are visible indicating the coupling between the membrane and the cytosol fluid. The shapes of the RBC under loading conditions agree with ones from experimental data of Mills et al. \cite{Mills2004}. The computed values of the axial ($D_a$) and transverse ($D_t$) diameters agree well with the experimental values as seen in Figure \ref{fig:streching_test}a. In particular, the values of $D_a$ and $D_t$ are consistent across the different values of $N_v$, which indicate a robust performane of the coarse-graining procedure. There is a disagreement between the simulated results and the experimental value of $D_t$. Examining the shapes of the RBC in the simulations (\ref{fig:streching_test}b), it is revealed that the RBC tends to rotate around the stretching direction. This rotation leads to the difference between the experimental and numerical results of $D_t$. In brief, the mechanics of RBC is well replicated by the computational model across different level of coarse-graining. Thus, a value of $N_v= 1000$ is chosen to report the dynamics of the RBC in subsequent sections.

\subsubsection{Deformation of the RBC under constant shear rates $\overline{\dot{\gamma_0}}$} \label{sec:RBCExtracellularflow}
\begin{figure}[H]
    \centering
    \includegraphics[width=.75\linewidth]{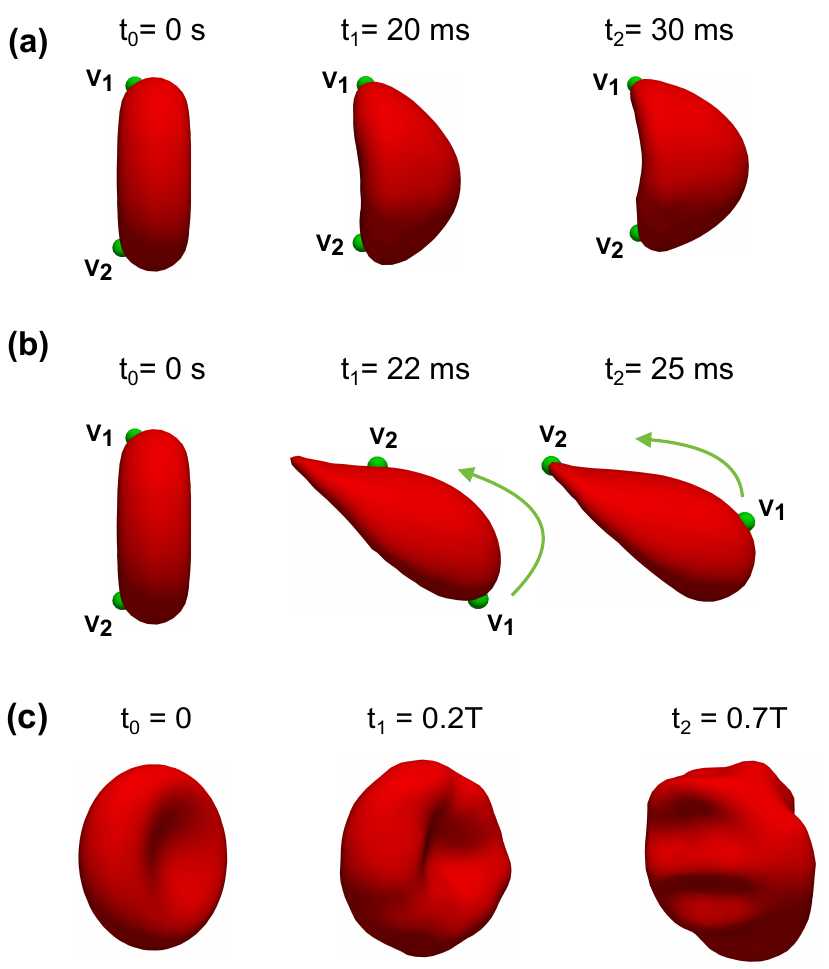}
    \caption{The transitions from the idealized shape to realistic shapes under the impact of shear flows. The stable shapes are attained under the impact of constant shear rate ($I_0$) in: $(a)$ croissant shape $(I_0 U_3 r_1 \chi_1)$ and $(b)$ slipper shape $(I_0 U_4 r_3 \chi_1)$. The RBC membrane only exhibits the tank-treading effect in the slipper shape $(I_0 U_4 r_3 \chi_1)$, which is characterized by the motions of two Lagrangian markers $V_1$ and $V_2$. The slipper shape is maintained by the counter-clockwise rotation (the green arrow) of the cellular membrane around the RBC's centroid. The multilobe shape appears $(c)$ under the oscillatory flow ($I_s\psi_6r_1\chi_3$) during the backward phase ($0.7T$).}
    \label{fig:tank_treading_evolution}
\end{figure}

Under constant shear rate conditions ($I_0$) as described in section \ref{sec:constant_shear_rate}, two districts of the RBC shape are observed: $(i)$ the croissant shape ($I_0U_3r_1 \chi_1$ - $\overline{\dot{\gamma_0}}=200~s^{-1}$); and $(ii)$ the slipper shape ($I_0U_4r_3\chi_1$ - $\overline{\dot{\gamma_0}}=600~s^{-1}$) as shown in Figure \ref{fig:tank_treading_evolution}. 

Under low shear rate ($I_0U_3r_1\chi_1$), the RBC was initially placed along the centerline of the microchannel (discocyte shape). As the RBC interacts with the incoming flow,  deforms, and eventually transitions to a croissant shape. The terminal shape (croissant) is attained as the RBC continues to propagate along the channel's symmetry axis as shown in Figure \ref{fig:tank_treading_evolution}a. Note that the croissant shape in this case is not fully axi-symmetric as the RBC is immersed in a rectangular channel. 

Under high shear rate ($I_0U_4r_3\chi_1$), the RBC transitions from the croissant shape to the slipper shape as shown in Figure \ref{fig:tank_treading_evolution}b, which exhibits a bistability mode with tank-treading behavior.  Note that the RBC is placed at a radial shift $r_3 = 0.7~\mu m$. Thus the initial location of the RBC is not at the channel's symmetry axis. The tank-treading effect is a complex dynamics in which the RBC membrane propagates axially along the channel while it rotates around its own center of mass. This rotation of the membrane/cytoskeleton around the cytoplasm is shown clearly in Figure \ref{fig:tank_treading_evolution}b. A counter-clockwise rotation is observed as indicated by the locations of two membrane particles (Lagrangian points - $V_1$ and $V_2$) at different time instances ($t_1=22~ms$ and $t_2=25~ms$).

In both the croissant or slipper shapes, the shape transition from the initial shape (discocyte) to the terminal shape (either croissant or slipper) occurs within around $30~ms$. These transitions agree well with the corresponding experimental data of Guckenberger et al. (2018)\citep{Guckenberger2018} as well as described in recent experiments on RBC transient dynamics \cite{recktenwald2022red, prado2015viscoelastic}. Furthermore, our shapes (croissant and slipper) for confined flow are in good agreement with the shape diagram produced by Agarwal et al. \cite{Agarwal2022} for different Capillary numbers and confinements as seen in Figure \ref{fig:shape_diagram_validation}. In conclusion, our simulations are able to replicate the dynamics of the croissant and slipper shapes excellently well.

\begin{figure}
    \centering
    \includegraphics[width=.8\linewidth]{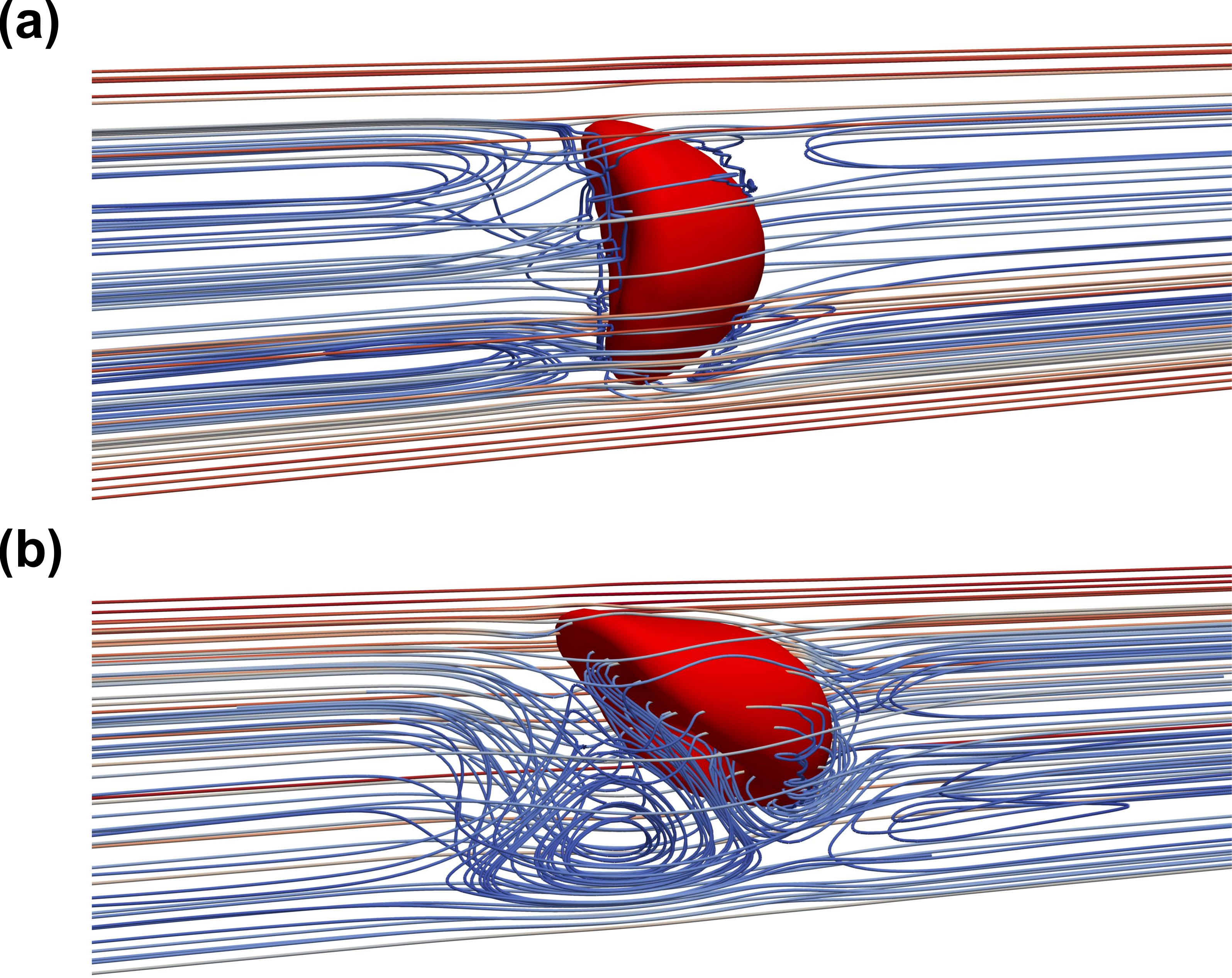}
    \caption{The extracellular flow patterns for: ($a$) the croissant shape ($I_0U_3r_1\chi_1$)  and ($b$) the slipper shape ($I_0U_4r_3\chi_1$). The flow streamlines are reconstructed using the co-moving frame method as discussed in section \ref{sec:RBCExtracellularflow}. The tank-treading effect induces a closed vortex to form on the upstream side of the RBC.}   \label{fig:croissant_transition_streamlines}
\end{figure}

The extracellular patterns of the croissant and slipper shapes agree excellently well with the experimental data of Guckenberger et al. (2018) \cite{Guckenberger2018}. The extracellular flow pattern can be visualized by reconstructing the relative flow velocity field \citep{Yaya2021}. The relative velocity is defined is the difference between the flow velocity and the RBC's centroid velocity as shown in Figure \ref{fig:croissant_transition_streamlines}. In the croissant shape ($I_0U_3r_1\chi_1$), the velocity streamlines closely resemble an axi-symmetrical flow pattern (Figure \ref{fig:croissant_transition_streamlines}a). The downstream side of the RBC membrane deforms significantly whereas the upstream side barely changes as depicted in Figure \ref{fig:croissant_transition_streamlines}b. In the slipper shape ($I_0U_4r_3\chi_1$), there exists an asymmetrical vortical structure in the vicinity of the RBC membrane. As the slipper shape emerges, a fully closed vortex ring is created by a reversed flow region, which is close to the channel wall. In short, the emergence of the RBC shape dictates the extracellular flow pattern.

\subsubsection{Propulsion of RBC under stepwise oscillatory flows ($I_s$)}
Under stepwise flow waveform ($I_s$), our simulation results agree well with the propulsion step map ($\Delta x_c, Ca^f$), which was developed by Schmidt et al. (2022) \cite{schmidt2022oscillating} for both channels $\chi_2 = 0.5$ and $\chi_3 = 0.38$.  In both cases, the propulsion step $(\Delta x_c)$ was observed to monotonically increase with the values of $Ca^f$. However, the $\Delta x_c$ is higher in the lower confinement channel ($\chi_3$), which indicates the importance of channel confinement. In all simulation cases $(I_s \psi_1 r_1 \chi_3)$, $(I_s \psi_2 r_1 \chi_3)$, $(I_s \psi_3 r_1 \chi_3)$, $(I_s \psi_4 r_1 \chi_3)$, $(I_s \psi_5 r_1 \chi_3)$ and $(I_s \psi_6 r_1 \chi_3)$ the RBC transitioned from the discocyte to the biconcave shape during the forward phase ($0 < t < \frac{T}{5}$) with all values of the peak forward flow ($\psi_f = 1.05~mm/s$ to $\psi_f = 4.34~mm/s$) as shown in Figure \ref{fig:tank_treading_evolution}c. Strikingly, the complex multilobe shape emerges during the backward phase $T_b$. The elastic response of the RBC membrane to the oscillatory flow during the cycle $T$ is depicted for the case $(I_s \psi_6 r_1 \chi_3)$ in Figure \ref{fig:tank_treading_evolution}c. The reversal of the flow direction during $T_b$ results in membrane buckling and stretching, which give rise to the multilobe shape even if the RBC is placed initially at the channel center ($r = r_1 = 0$).

\subsection{The impact of oscillatory flows on RBC dynamics}
\subsubsection{The emergence of RBC shapes}
\begin{figure}[H]
    \centering
    \includegraphics[width=0.8\linewidth]{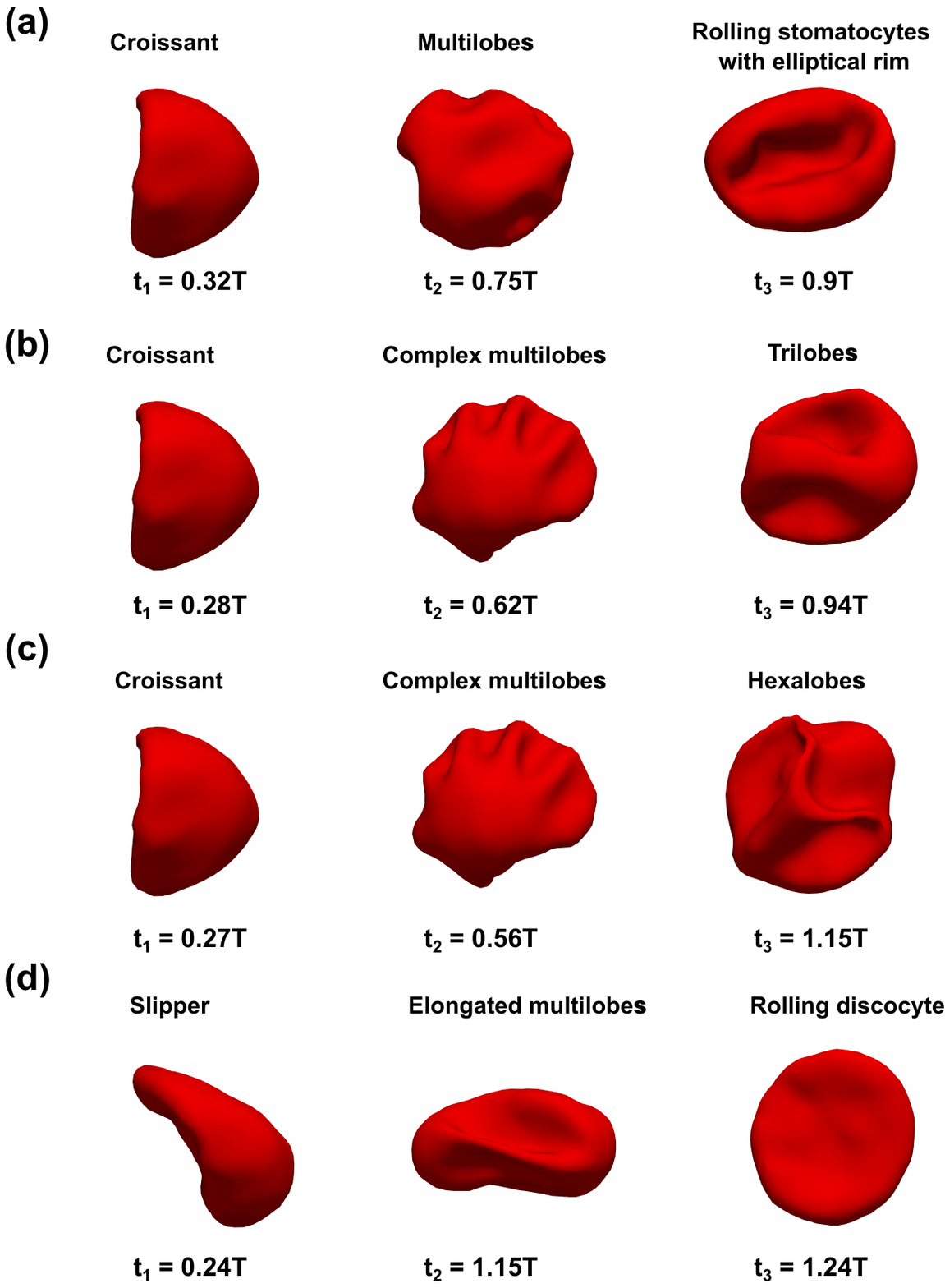}
    \caption{The emergence of complex shapes induced by different inlet sinusoidal waveforms. $(a)$ $I_1U_1r_1\chi_1$, $(b)$ $I_3U_2r_1\chi_1$, $(c)$ $I_4U_2r_1\chi_1$ and $(d)$ $I_4U_3r_3\chi_1$.}
    \label{fig:RBC_inflow_shapes}
\end{figure}

The oscillatory flow waveform ($U(t)$) further adds complexity to the membrane dynamics as the shape of RBC is highly sensitive to the extracellular flow condition. As the result of the pulsatile flow condition, the RBC shape continuously responds to the applied flow in the channel. Our simulations show that the RBC alternates its shapes in one of the following types: 1) croissant; 2) slipper; 3)trilobes; 4)  simple/complex/elongated multilobes; 5) rolling stomatocytes; 6)hexalobes; and 7) rolling discocyte as shown in Figure \ref{fig:RBC_inflow_shapes} and Tables \ref{tab:I1}-\ref{tab:I4}. The emergence of each type will be discussed as follows.

In all cases, the RBC evolves from the croissant ($C$) toward the slipper ($S$) mode during the forward phase ($0 < t < T_f$) of the flow cycle ($\frac{t}{T} \approx 0.25$) as shown in Figure \ref{fig:RBC_inflow_shapes}. Note that the transition to $C$ or $S$ mode from the biconcave shape is dependent on the value of the radial shift $(r)$. As shown in Tables \ref{tab:I1}-\ref{tab:I4}, the $S$ mode appears only when the RBC is initially placed  not exactly at the cross-sectional center ($r > 0$). The RBC remains in $C$ mode during the forward phase if it is initially placed at the cross-section center ($r = 0$) regardless of the bulk flow waveform ($I_1$ to $I_4$). In brief, the croissant and the slipper shapes exists during the forward phase and their emergence depends on the initial off-centered location of the RBC ($r$).

The RBC transitions from the simple shapes (croissant and slipper) toward more complex shapes (trilobes, simple/complex/elongated multilobes, rolling stomatocytes, hexalobes, and rolling discocyte) later in the flow cycle during the resting/reverse periods ($\frac{t}{T} > 0.5$). The shape transformation is initiated by the buckling of the RBC membrane, which takes place in the resting interval ($T_r$) phase of the flow (see Figure \ref{fig:waveform_types}). As a result of the change in flow direction, the RBC experiences considerable stretching and compression, leading to significant alterations in its membrane shape. 

\subsubsection{The impacts of the initial position $(r)$ and waveform $(I)$}
Our finding (Figure \ref{fig:RBC_inflow_shapes}) revealed clearly that the initial position $(r)$ and the flow waveform $(I)$ play a critical role in the emergence of RBC shapes. Under the symmetric and asymmetric waveforms, the RBC placed initially at the channel axis $r = r_1 = 0$, transitions sequentially from the croissant shape toward the complex multilobe, multilobe, trilobe, rolling stomatocyte, elongated multilobe, and finally hexalobe as shown in the Tables \ref{tab:I1}-\ref{tab:I4}. When $r > 0$, the RBC remains mostly the slipper shape during the forward phase ($t < T_f$) and it transitions toward the elongated multilobe during the back flow phase ($ T_f + T_r < t < T$). Finally, the RBC becomes a rolling discocyte in the second cycle ($t \approx 1.2  T$). In brief, the shape transition process is strongly sensitive to the initial placement of the RBC.

\begin{figure}[H]
  \vspace{-2cm}
    \hspace{-1cm}
\includegraphics[width=1.0\linewidth]{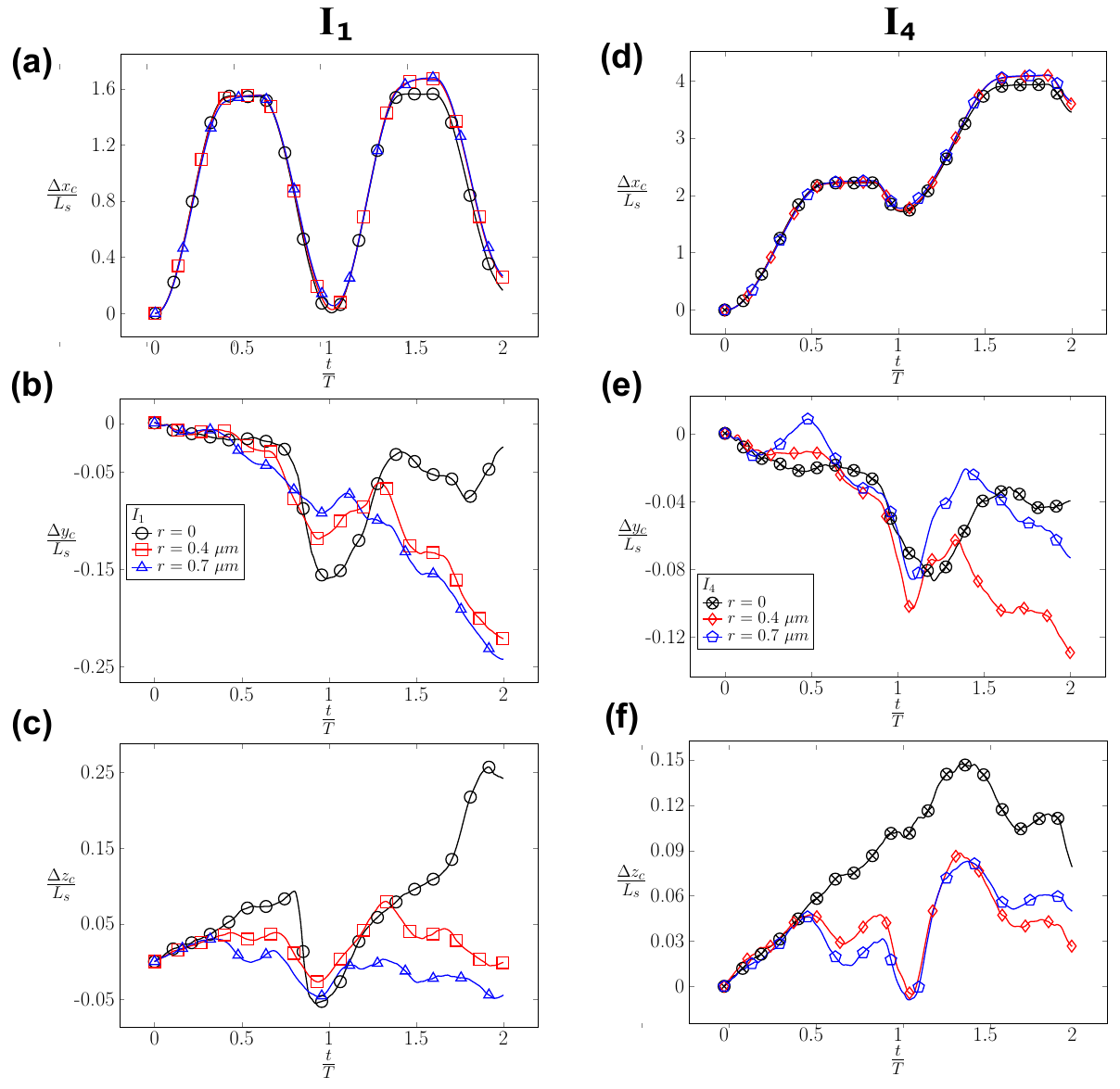}
    \caption{The impacts of the initial position $(r)$ on the time evolution of the RBC's centroid displacement ($\Delta x_c, \Delta y_c, \Delta z_c$). The instantaneous evolution of the RBC’s centroid position ($x_c$, $y_c$, $z_c)(t)$ is recorded as the RBC propagates along the channel. The displacements of the RBC from its initial location along three directions ($\Delta x_c$, $\Delta y_c$, $\Delta z_c$) are measured in units of the length scale $L_s$. The evolution of the centroid position is examined under two conditions: $(i)$ the symmetric $I_1$ (left column- $(a-c)$ ); and  $(ii)$ the asymmetric $I_4$ (right column - $(d-f)$) waveforms at different values of the radial shift $r_1, r_2$, and $r_3$. The symmetrical flow cases (left column) include $I_1 U_1 r_1\chi_1$, $I_1 U_1 r_2\chi_1$, and $I_1 U_1 r_3\chi_1$. The asymmetrical cases (right column) include $I_4 U_1 r_1\chi_1$, $I_4 U_1 r_2\chi_1$, and $I_4 U_1 r_3\chi_1$ ) cases. The exact values of $r_1$, $r_2$, and $r_3$ are described in Table \ref{tab:summary_factors}.}
    \label{fig:I_1_4U_1r_x_y_z}
\end{figure}

It is striking to observe the irreversible dynamics of RBC. When subjected to symmetric waveform $(I_1)$, the RBC is observed to be fully controlled by the pulsatile inflow. The RBC oscillates around its initial position with a minimal propulsion. Despite the inflow waveform is completely symmetrical (a sine function - $I_1$), the axial position of the RBC in Figure \ref{fig:I_1_4U_1r_x_y_z}a (left column) shows a positive value of the displacement $\Delta x_c$ at the end of the first ($t=T$) and second cycle ($t = 2T$) even when there is no radial shift ($r = 0$). Though small, this positive value of $\Delta x_c$  indicates that the RBC does not go back exactly to its initial location, which is $\Delta x_c = 0$ at $t = 0$. At all values of the radial shift of $r = 0, 0.4$, and $0.7 \mu m$, this irreversible dynamics is even more evident as shown in the lateral displacements in Figures \ref{fig:I_1_4U_1r_x_y_z}b-c. The magnitudes of $\Delta y_c$ and $\Delta z_c$ are comparable for all values of $r$ during the cycles.  For the case $I_1 U_1 r_1 \chi_1$ $(r = 0)$ the value of $\Delta y_c$ reaches a value of approximately $0.16 L_s$ at the end of the first cycle. For the cases $I_1 U_1 r_2 \chi_1$ and $I_1 U_1 r_3 \chi_1$, the values of $\Delta y_c$ and $\Delta z_c$ reach approximately $0.25 L_s$ at the end of the second cycle. In the vertical direction ($z_c$) in Figure \ref{fig:I_1_4U_1r_x_y_z}c, the well-centered RBC ($r=0$) was influenced by the change of flow direction, which is depicted by the upward and downward trends in the first cycle. However, the cell followed a dominant upward trend during the entire second cycle resulting in a lateral migration of around $0.25L_s$. Therefore, there exist significant lateral migration of the RBC during its propagation regardless of its initial position in the symmetrical waveform condition ($I_1$). In conclusion, a symmetrical flow waveform ($I_1$) results in a minimal propulsion along the axial direction but a significant lateral migration.

Under asymmetric waveform $I_4$, the RBC propels along the channel direction with a propulsion step of approximately $2L_s$ in each cycle as shown in Figure \ref{fig:I_1_4U_1r_x_y_z}d. As the waveform becomes asymmetric with a longer forward phase, the RBC does not go back significantly during the reverse phase. It rather remains at the displacement value of $\Delta x_c \approx 1.9 L_s$ at the end of the first cycle. It continues to propel in the second cycle up to $\Delta x_c \approx 4.0 L_s$. Surprisingly, the lateral migration of the RBC ($\Delta y_c, \Delta z_c$) is smaller in comparison to ones in the symmetric case ($I_1$). The values of ($\Delta y_c, \Delta z_c$) are within $0.15 L_s$ for all cases $I_4 U_1 r_1 \chi_1$, $I_4 U_1 r_2 \chi_1$, and $I_4 U_1 r_3 \chi_1$ as shown in Figure \ref{fig:I_1_4U_1r_x_y_z}e and f. In brief, the RBC propels significantly under the impact of the asymmetrical flow waveform $I_4$ along the axial direction but it does not migrate significantly in the lateral directions.

\begin{figure}[H]
    \centering
    \vspace{-2cm}
    \includegraphics[width=1\linewidth]{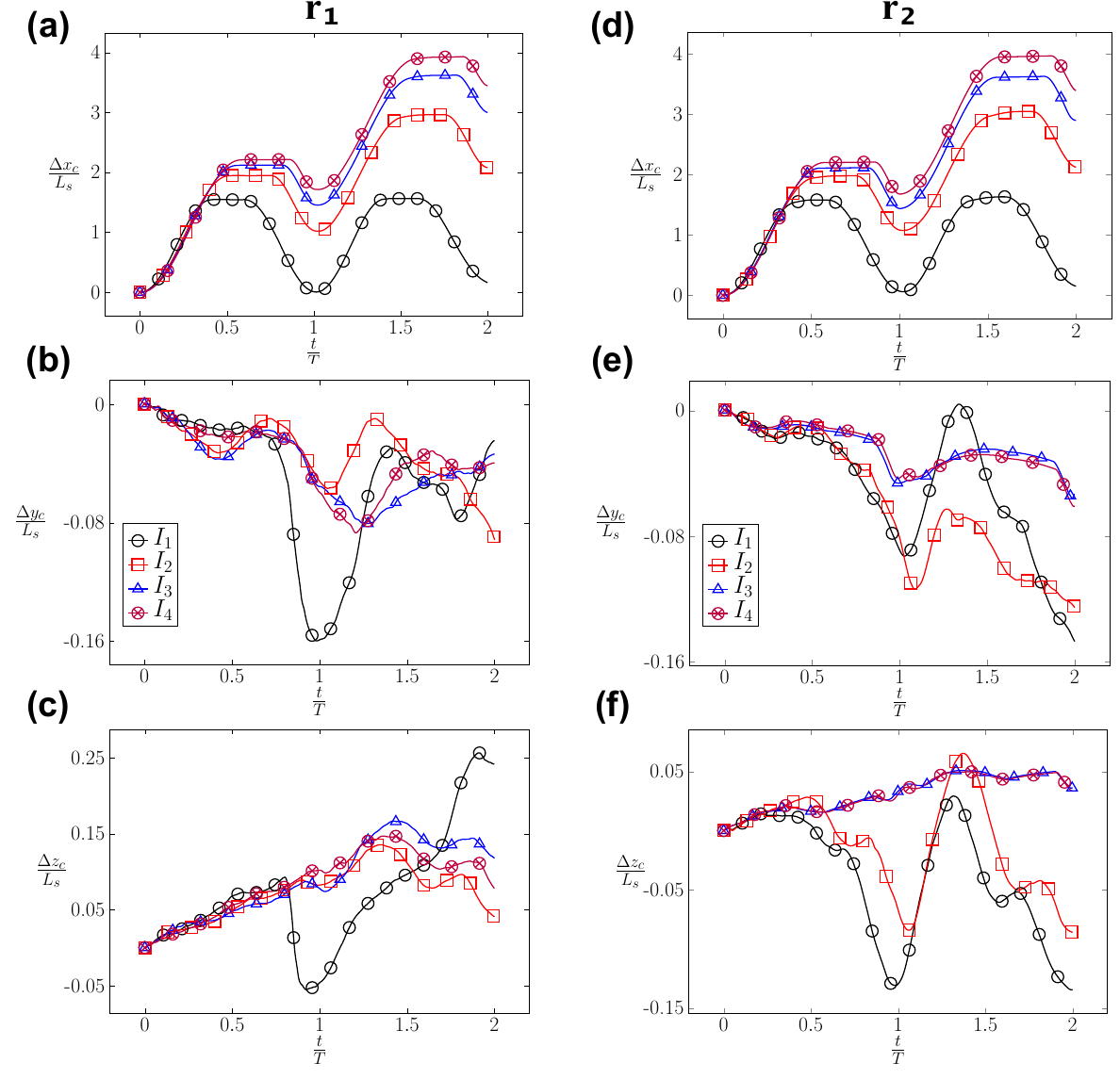}
    \caption{The impacts of the waveform $(I)$ on the time evolution of the RBC's centroid displacement ($\Delta x_c, \Delta y_c, \Delta z_c$). The instantaneous evolution of the RBC’s centroid position ($x_c$, $y_c$, $z_c)(t)$ is recorded as the RBC propagates along the channel. The displacements of the RBC from its initial location along three directions ($\Delta x_c$, $\Delta y_c$, $\Delta z_c$) are measured in units of the length scale $L_s$. The evolution of the centroid position is examined under two conditions: $(i)$ centred initial position $r = r_1 = 0$ (left column- $(a-c)$) for the cases $I_i U_1 r_1\chi_1$; and  $(ii)$ off-centered initial position $r = r_2 = 0.4~\mu m$ (right column - $(d-f)$) for the cases $I_i U_2 r_2\chi_1$ with $i = 1, 2, 3$, and $4$ as described in Table \ref{tab:summary_factors}.}
\label{fig:I1_I4_r0_X_Y_Z_plots_Combined}
\end{figure}

When the RBC is positioned at the center line $(r = 0)$ of the channel, it is observed to be fully controlled by the pulsatile inflow when subjected to a symmetric waveform ($I_1$) as shown in Figure \ref{fig:I1_I4_r0_X_Y_Z_plots_Combined}a. In this case, the cell oscillates around its initial position with minimal propulsion. However, as the inflow profile transitions to asymmetric waveform ($I_2$, $I_3$, and $I_4$) with an increasing forward velocity time interval, the RBC gains more momentum and propels far away from its initial position reaching a maximum propulsion step $\Delta x_c$ of approximately $4L_s$ at the end of the second cycle. In the lateral direction $(y_c)$, $\Delta y_c$ reached a value of approximately $0.16 L_s$ at the end of the first cycle when subjected to symmetric waveform $(I_1)$ as shown in Figure \ref{fig:I1_I4_r0_X_Y_Z_plots_Combined}b, while for the cases $I_2$, $I_3$ and $I_4$ the values of $\Delta y_c$ were comparable at the end of the second cycle, especially as the waveform becomes predominantly asymmetric ($I_3$ and $I_4$). Furthermore, in the vertical direction $(z_c)$ as seen in Figure \ref{fig:I1_I4_r0_X_Y_Z_plots_Combined}c, the RBC follows a monotonically upward trend throughout the entire second cycle. This results in a vertical propulsion $\Delta z_c$ of approximately $0.25L_s$. However, for all the asymmetric waveforms, a nearly identical upward trend is observed, leading to a vertical displacement $\Delta z_c$ of about $0.08L_s$ at the end of the first cycle. However, during the entire second cycle, the cell is observed to oscillate with a downward trend. In summary, the symmetric waveform leads to the maximum lateral and vertical propulsion, while the asymmetric waveforms results in the maximum axial propulsion.

The off-centered $(r = 0.4~\mu m)$ axial migration of the RBC exhibited a behavior similar to the centered case, indicating that the initial position does not significantly affect the axial propulsion of the RBC. In the lateral direction, the RBC under $I_1$ and $I_2$ achieved a lateral propulsion of approximately $0.16L_s$ (Here $L_s = 8 \mu m$)at the end of the second cycle. While the centered case reached this value at the end of the first cycle, the off-centered initial placement resulted in a slower lateral propulsion due to the cell experiencing a gradient of velocity magnitude compared to the centered case. Additionally, $I_3$ and $I_4$ displayed nearly identical profiles with a maximum propulsion of $0.06L_s$. A similar pattern was observed in the vertical direction, where the RBC under $I_1$ and $I_2$ exhibited similar oscillation profiles, reaching a propulsion step of approximately $0.14L_s$ at the end of both cycles. On the other hand, $I_3$ and $I_4$ displayed a nearly identical steady upward trend throughout the entire two cycles, resulting in a vertical propulsion of approximately $0.04L_s$. To summarize, the differentiation observed between $I_1$, $I_2$, and $I_3$, $I_4$ implies that when the RBC is off-centered, a maximum critical forward time interval is considered in order to attain the highest propulsion. Based on the findings of this study, to achieve maximum propulsion in all directions, the forward time interval $(T_f)$ should be less than three times the backward time interval $(T_b)$, expressed as $\frac{T_f}{T_b} < 3.$

\subsubsection{Extracellular flow dynamics at the vicinity of the RBC under oscillatory flows}
\begin{figure}[H]
    \hspace{-1cm}
    \centering
    \includegraphics[width=1.0\linewidth]{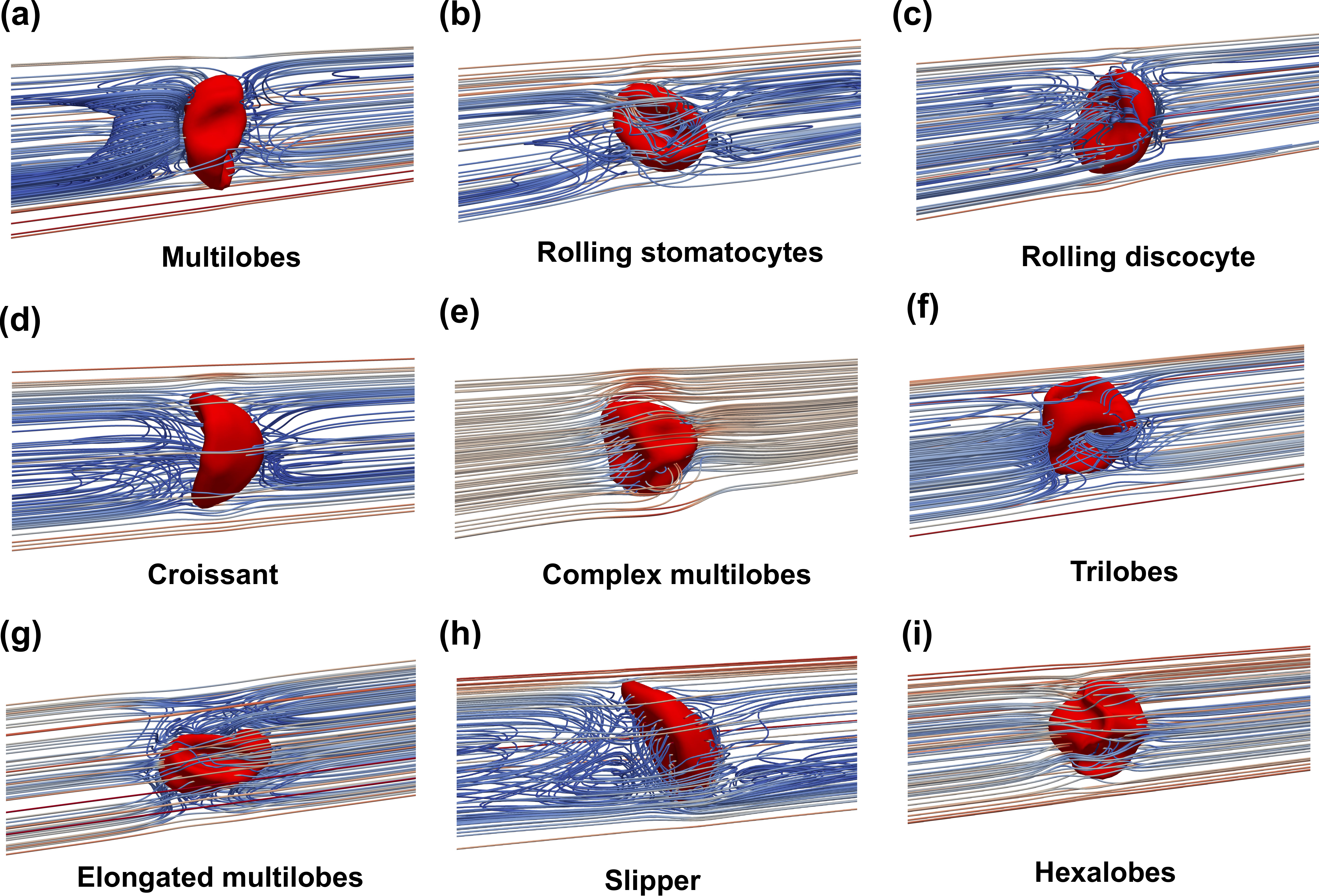}
   \caption{Snapshots of the 3D flow streamlines surrounding the RBC under different oscillatory flow conditions, which correspond to the observed shapes in Figure \ref{fig:RBC_inflow_shapes}. Here, the 3D oscillatory flow streamlines corresponding to the multilobes and rolling stomatocyte were examined in the case of $I_1U_1r_1\chi_1$, and rolling discocyte from the case $I_1U_1r_2\chi_1$. Furthermore, the streamlines for the croissant, complex multilobes, trilobes, and elongated multilobes shapes were visualized from the case of $I_1U_3r_1\chi_1$. Additionally, the streamlines for the slipper and hexalobes shapes were recorded from the cases $I_1U_3r_3\chi_1$ and $I_4U_2r_1\chi_1$, respectively.}
   \label{fig:RBC_oscillatory_streamlines}
\end{figure}

The emergence of the RBC shape has a close relationship with the flow pattern of the surrounding fluid (extracellular flow). Under the impact of the channel confinement, the deformation of RBC is well regulated by the flow waveform, which result in distinct extracellular flow patterns surrounding the RBC as shown in Figures \ref{fig:RBC_inflow_shapes} and \ref{fig:RBC_oscillatory_streamlines}. To highlight the  impact of the RBC motion, the  flow pattern is visualized in the co-moving frame with the RBC's centroid (see section \ref{sec:RBCExtracellularflow}). Thus, the flow streamlines are represented from the perspective of the RBC.

The case $(I_1U_1r_1\chi_1)$ is selected to illustrate the evolution of flow pattern as the RBC deforms from a relatively simple shape to a more complicated shape as depicted in Figure \ref{fig:RBC_oscillatory_streamlines} (first row). This case is chosen because the temporal variation of the waveform is completely symmetrical ($I_1$). Moreover, the RBC is placed initially at the channel axis ($r = r_1 = 0$) with the lowest forward velocity $\psi_f = U_1 = 1~mm/s$. In the case $(I_1U_1r_1\chi_1)$, Figure \ref{fig:RBC_oscillatory_streamlines}a revealed that the RBC has a multilobe shape at the end of the forward phase. The presence of the large lobes resulted in a more convoluted streamline patterns during the resting phase. As the RBC undergoes a morphological transition to rolling stomatocye at the end of the first cycle $(t = 0.9T)$, the streamlines experienced changes (Figure \ref{fig:RBC_oscillatory_streamlines}b). However, when the RBC transformed into rolling discocyte in the case $I_1U_1r_2\chi_1$ shown in Figure \ref{fig:RBC_oscillatory_streamlines}c, the streamlines once again resembled to similar patterns observed in the croissant shape (constant shear rate case $I_0U_3r_1\chi_1$ in Figure \ref{fig:croissant_transition_streamlines}a). 

The case $(I_1U_3r_1\chi_1)$ is selected to illustrate further the impact of the peak forward flow $\psi_f$. In this case, the peak velocity $\psi_f$ is increased to $\psi_f = U_3 = 2~mm/s$ while other parameters are kept unchanged in comparison to $I_1U_1r_1\chi_1$. Therefore, the most significant factor of the shape transition is due to the impact of the peak inlet velocity $\psi_f = U_3 = 2~mm/s$. The RBC transitions quickly to the croissant shape in Figure \ref{fig:RBC_oscillatory_streamlines}d $(t = 0.28T)$. The flow patterns are similar to those observed under constant shear rate  (see case $I_0U_3r_1\chi_1$ in Figure \ref{fig:croissant_transition_streamlines}a). 
During the rest period ($ T_f < T < T_f + T_r$), the flow velocity surrounding the cell decreased notably and the complex multilobes shape emerges as seen in Figure \ref{fig:RBC_oscillatory_streamlines}e. The flow pattern is perturbed minimally surrounding the RBC as it shape turns to trilobe in Figure \ref{fig:RBC_oscillatory_streamlines}f.  During the backward phase $(t = 1.15 T)$, the RBC becomes further elongated as its lobes are stretched further. Consequently, the flow patterns in the vicinity of the cell exhibit pronounce transience as shown in Figure \ref{fig:RBC_oscillatory_streamlines}g. In brief, the peak velocity $\psi_f$ can induce complex morphology of the cell as well as the associated surrounding fluid flows.

To highlight the impact of the initial location $r$, the case $I_1U_3r_3\chi_1$ was selected to visualize the flow patterns. As shown in Figure \ref{fig:RBC_oscillatory_streamlines}h, due to the off-centered initial location ($r > 0$) the slipper shape emerges during the forward phase. A closed vortex ring is also observed downstream of the RBC as the flow velocity reaches its maximum magnitude in the forward phase. This phenomenon is similar to the one observed in the constant shear rate case ($I_0U_4r_3\chi_1$ with $U_4 = 6~mm/s$) in Figure \ref{fig:croissant_transition_streamlines}b. This is remarkable since the peak flow $\psi_f$ is rather three times lower in this case $\psi_f = U_3 = 2 ~ mm/s$.

Furthermore, the hexalobes shape (observed only in the case $I_4U_2r_1\chi_1$) corresponding flow patterns are shown in Figure \ref{fig:RBC_oscillatory_streamlines}i. During the resting period $(t = 1.15T)$, the extracellular flow exhibits a minimal disturbance around the hexalobes as the RBC completed the transition in the rest period.

\section{Discussion}
Due to the membrane flexibility, RBC responds swiftly to the applied shear rate \citep{Lanotte2016}. This characteristic can be exploited to understand the mechanical properties of the RBC membrane \cite{prado2015viscoelastic} and thus it has the potentials to identify the pathological changes \cite{recktenwald2022red} of RBC's membrane. However, the exact mechanism of this response are not yet fully understood. In this work, we explore the impacts of the unsteady shear rate to control cell deformation and migration in micro-channels.

Our numerical method is based on the concept of coupling continuum-particle methods \citep{akerkouch2021}, which allows the simulations of RBC dynamics under physiological conditions. Our numerical results showed an excellent agreements with available $in~ vitro$ and computational studies both in cellular mechanics and extracellular flow pattern of the blood plasma\citep{Mills2004, Guckenberger2018, Yaya2021}. While most previous studies \cite{Fedosov2010, Pivkin2008} have only focused on the impact of constant shear rate on the dynamics of the RBCs, our results show that the unsteady shear rate can induce complex RBC's morphology as discussed below. 

\subsection{The emergence of the croissant shape and the slipper shape under a constant shear rate $\overline{\dot{\gamma_0}}$)}
\begin{figure}[H]
    \centering
    \includegraphics[width=0.8\linewidth]{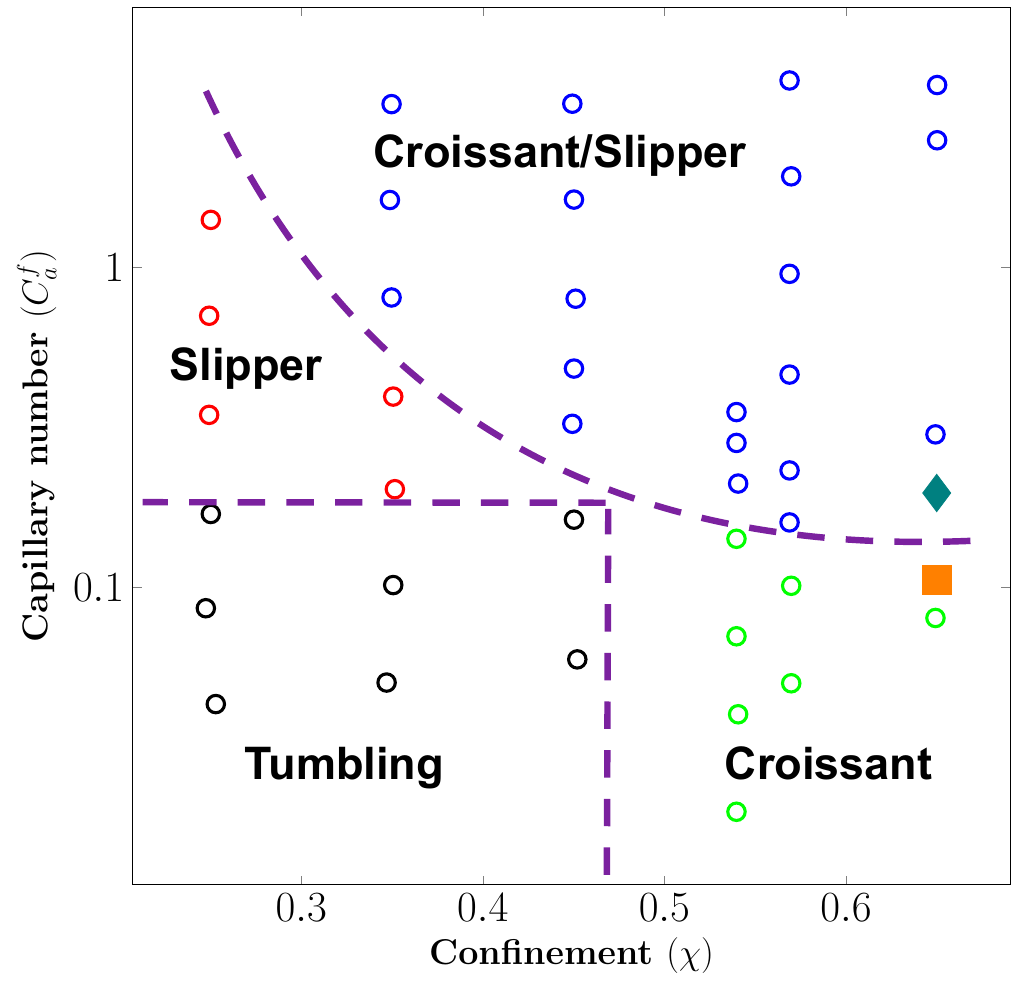}
    \caption{Validation with the shape diagram ($Ca$, $\chi$) of Agarwal et al. (2022) \cite{Agarwal2022} (unfilled circles ) for flows of RBC in a confined channel. The dash lines depict distinct regions representing different dynamics/shapes of RBC. Our simulations for the croissant shape ($I_0 U_3 r_1 \chi_1$ - filled square) and the slipper shape ($I_0 U_4 r_1 \chi_1$ - filled diamond) agree well with the reported regions with $\chi = \chi_1 = 0.65$}.
    \label{fig:shape_diagram_validation}
\end{figure}

In micro-channel flows with constant shear rate ($\overline{\dot{\gamma_0}}$), three common dynamics of RBCs are frequently observed: (i) tumbling; (ii) croissant/parachute; and (iii) slipper shapes as shown in Figure \ref{fig:shape_diagram_validation}. In unconfined flows \cite{dupire2012full}, the RBC dynamics depends on only the shear rate ($\dot{\gamma}$ or the $Ca$) and viscosity contrast ($\lambda$). However, the confinement of micro-channel flows imposes an additional condition for shape transition via the confinement ratio $\chi$. As shown in Figure \ref{fig:shape_diagram_validation}, the combination of $Ca$ and $\chi$ dictates to the RBC shape either the croissant or slipper shapes. 

Recent works \cite{Guckenberger2018, Yaya2021} in rectangular microchannels, which are identical to our channels as shown in Figure \ref{fig:simulation_setup} and Table \ref{tab:Channels}, further suggest that the emergence of RBC shape is also dependent on the radial shift ($r$ - see Figure \ref{fig:simulation_setup} for its definition). On one hand, the croissant shape dominates when the RBC is placed initially at the cross-sectional center with large confinement. In previous works \cite{Guckenberger2018, Yaya2021}, the croissant shape emerged at low shear rate $(\overline{\dot{\gamma_0}} < 300~s^{-1})$ if the RBC is placed exactly at the channel's center ($r=0$). On the other hand, the slipper shape emerge whenever the RBC was not placed exactly at the centerline ($r > 0$). The RBC was found to exhibit a (tank-treading) slipper shape at sufficiently high shear rate $(\overline{\dot{\gamma_0}} \approx 500~s^{-1})$ and off-centered placement ($r > 0$) \cite{Guckenberger2018, Yaya2021}. In cylindrical micro-channels \cite{Fedosov2014}, similar observations were confirmed albeit at lower shear rates $(0 < \overline{\dot{\gamma_0}} < 80~s^{-1})$. Therefore, the radial shift plays an important role in RBC dynamics.

Our results in Figure \ref{fig:shape_diagram_validation} confirm the croissant-to-slipper transition as the Capillary number (and thus $\overline{\dot{\gamma_0}}$) increases from 0.1 to 0.37 for a confinement of $\chi = 0.65$. The croissant shape emerges when the initial position of the RBC is placed exactly at the channel centerline at sufficiently low shear rate ($Ca = 0.1$). When the shear rate is increased to $Ca = 0.37$, the slipper shape emerges. Furthermore, our model is able to capture the intricate dynamics of the tank-treading motion, which is characterized by the rotation of the membrane at the shear rate of $600~s^{-1}$ as illustrated in Figure \ref{fig:tank_treading_evolution}. Therefore, our results further confirm the importance of the radial shift.
\subsection{The impact of time-varying shear rate $\overline{\dot{\gamma(t)}}$ on RBC shape }
When the inflow varies in a stepwise manner as seen in Figure \ref{fig:cfd_validation}, the shear rate ($\bar{\dot{\gamma}}$) changes as a function of time $\bar{\dot{\gamma}}(t)$ with distinct forward ($T_f$) and backward ($T_b$) time phases. In all cases ($I_s \psi_1 r_1 \chi_2$, $I_s \psi_2 r_1 \chi_2$, $I_s \psi_3 r_1 \chi_2$, $I_s \psi_4 r_1 \chi_3$, $I_s \psi_5 r_1 \chi_3$, and $I_s \psi_6 r_1 \chi_3$), the RBC is placed exactly at the channel axis $(r = r_1 = 0)$. The RBC transitions from a discocyte shape toward the croissant shape during its propulsion as shown in Figure \ref{fig:tank_treading_evolution}. Although the backward phase induces the buckling of the cellular membrane, the RBC shape remains symmetrical with respect to the channel axis (multilobes) as shown in Figure \ref{fig:tank_treading_evolution} at the end of $T_b$. This is remarkable given that the maximum shear rate during the backward phase can be sufficiently large ($\overline{\dot{\gamma}}_f = 207~s^{-1}$). 
Comparing the case $I_0 U_3 r_1 \chi_1$ and $I_0 U_4 r_3 \chi_1$ in Table \ref{tab:all_our_oscillaotry_cases}, our results suggest that the break of symmetry (croissant-to-slipper transition \cite{recktenwald2022red}) is observed only when the radial shift exists ($r > 0$). 

When applying different sinusoidal waveforms $(I_1$, $I_2$, $I_3$ and $I_4$) shown in Figure \ref{fig:waveform_types}, our results show the ubiquitous presence of croissant-to-slipper transition across all shear rates ($\bar{\dot{\gamma}}_f = 100, 150$, and $200~s^{-1}$). While the applied shear rate $\bar{\dot{\gamma}}(t)$ varies greatly over one cycle, the slipper shape appeared ($t \approx 0.3 T$) whenever the RBC is placed off the channel's axis ($r > 0$) as shown in Tables \ref{tab:I1}-\ref{tab:I4}. Note that these waveforms are different in term of the forward ($T_f$) and backward ($T_b$) phases, with the backward phase being the shortest in $I_4$. This explains the emergence of the slipper shape even when the waveform is reversible ($I_1$): $I_1 U_1 r_2 \chi_1$, $I_1 U_1 r_3 \chi_1$, $I_1 U_2 r_2 \chi_1$, $I_1 U_2 r_3 \chi_1$, $I_1 U_3 r_2 \chi_1$, $I_1 U_3 r_3 \chi_1$ . Hence our results indicate that the initial of the RBC in flows plays an essential role in determining the RBC dynamics. 

Our observations in Figure \ref{fig:RBC_inflow_shapes} and Tables \ref{tab:I1}-\ref{tab:I4} suggest that the shape transitions under reversible waveforms are accomplished through a consistent transient stretching and compression of the membrane. This occurs as the RBC experiences forward and backward flow phases during the cycle. Moreover, the orientation of the RBC's symmetry axis continuously changes relative to the symmetry axis of the channels. This suggests that the RBC moves in different directions depending on the initial conditions ($I$, $\bar{\dot{\gamma}}_f$ and $r$). 

In particular, experiments and numerical simulations using shear flows showed that the RBC under weak shear rates $( \bar{\dot{\gamma}} < 10~s^{-1})$ typically maintain its discocyte shape with an $80\%$ probability \cite{Lanotte2016}. However, as the shear rate gradually rises from $10~s^{-1}$ to $400~s^{-1}$, the likelihood of a discocyte shape decreases to $30\%$. The findings from Lanotte et al. \cite{Lanotte2016} demonstrate that the presence of the discocyte shape is correlated with weak shear rates. Their results have been found to hold true even when considering different viscosity ratios, as evidenced by the work of Mauer et al. \cite{Mauer2018} Our study consistently observed the discocyte shape during the second cycle, across all applied waveforms and shear rates ($\bar{\dot{\gamma}}_f = 100~s^{-1}$, $150~s^{-1}$, and $200~s^{-1}$) when the initial positions were off-centered, as indicated in Tables \ref{tab:I1}-\ref{tab:I4}.

Moreover, our findings in Figures \ref{fig:I_1_4U_1r_x_y_z} and \ref{fig:I1_I4_r0_X_Y_Z_plots_Combined} indicates that by the end of the first cycle the RBC underwent sufficient lateral propulsion in addition to the initial off-centered shift. This propelled movement led for the RBC to experience even lower shear rates closer to the channel's walls, facilitating the transition to the discocyte shape. However, under shear flow stomatocyte shape was observed to dominate the RBC population with $65\%$ when the shear rate is between  $(10~s^{-1} < \bar{\dot{\gamma}} < 400~s^{-1})$ \cite{Lanotte2016}, while we observed the elliptical-rim-shaped stomatocyte  only under symmetric waveform $I_1$ and centered initial placement $(r = r_1 = 0)$ subject for the shear rate of $100~s^{-1}$ $(I_1U_1r_1\chi_1)$. This results strongly suggest that the impact of waveform is significant in defining the morphology sequence the RBC can follow even at low oscillatory shear rates.

Our results underscore the significant influence of the applied waveform in shaping the morphological response of the RBC. At high constant shear rates $(400~s^{-1} < \bar{\dot{\gamma_0}} < 2,000~s^{-1})$ \cite{Lanotte2016}, polylobes shape emerges. This polylobes shape is characterized by large number of  lobes on the RBC surface, known as trilobes and hexalobes \cite{Lanotte2016}. The appearance of these polylobes is attributed to the substantial membrane buckling caused by the reverse of flow direction. In the current study, polylobes are also observed across all applied waveforms when the cell is placed initially at the channel axis ($r = 0$)  even at weak shear rates ($\bar{\dot{\gamma}}_f \leq 200~s^{-1}$) as in Figure \ref{fig:RBC_inflow_shapes} and Tables \ref{tab:I1}-\ref{tab:I4}. For example, the trilobes shape are observed in the reversible waveform ($I_1U_2r_1\chi_1$ and $I_1U_3r_1\chi_1$) or the irreversible waveform $(I_3U_2r_1\chi_1)$. Furthermore, the hexalobes shape only appears under the most reversible waveform with $(I_4U_2r_1\chi_1)$ ($r = 0$) and $\bar{\dot{\gamma_f}} = 150~s^{-1}$ as shown in Table XXXX. Surprisingly, we observed that the RBC can achieve this transition to polylobes over a short distance (approximately $4.0 \times L_s$ for $I_1U_2r_1\chi_1$ ) as shown in Figure \ref{fig:I1_I4_r0_X_Y_Z_plots_Combined}a.

The RBC shape can be further deformed into elongated shapes. Li et al. \cite{Li2014} demonstrated that as the shear rate increases, RBCs can undergo significant elongation and assume a more cylindrical shape. Our findings support this observation, as we also observed the elongated multilobes shape. Our results suggest that this shape is generally present regardless of the applied waveform, but it only manifests under higher shear rates. Specifically, we observed the elongated multilobes morphology for $\bar{\dot{\gamma}}_f \geq 150~s^{-1}$ under symmetric waveform and centered position, and $\bar{\dot{\gamma}}_f = 200~s^{-1}$ for all asymmetric waveforms and initial position.

\subsection{Controlling lateral migration of cells with oscillatory flows}
Microfluidic devices are typically used to isolate and separate cells \cite{gossett2010label} with different techniques. While these devices are promising for many cell-sorting applications \citep{Suresh2007, Brandao2003}, the main challenge is the difficulty in obtaining high-throughputs due to the required length of the microfluidic channels. Recent works have shown that varying the shear rates in time \citep{schmidt2022oscillating, Krauss2022} can reduce the required length based on the concept of velocity lift \cite{Qi2017}, which is the factor that drives the RBC's migration towards the center of the channel. 

As the inertial effect is negligible at very low Reynolds number ($R_e < 0.01$), the flow is reversible for a rigid body. Thus, a rigid body will return to its initial position if the inflow conditions in the backward phase is reversed in the exact opposite way of its own during the forward phase. However, the RBC is not a rigid body and its membrane is highly flexible. Our results in Figures \ref{fig:I_1_4U_1r_x_y_z} and \ref{fig:I1_I4_r0_X_Y_Z_plots_Combined} for the symmetrical waveform ($I_1$) show that the RBC does not go back to its initial position at the end of the cycle. There is an axial shift of the RBC from its original position ($\Delta x_c \neq 0$) at the end of the cycle. Moreover, the RBC migrates significantly in the lateral cross-section ($\Delta y_c \gg 0$ and ($\Delta z_c \gg 0$). Similar results were found experimentally \cite{Krauss2022} when the average positions of RBCs and stiff beads were compared in oscillatory flow. Thus, due to its soft nature the RBC showed a significant net actuation in asymmetric oscillating flows. This differential response points to the potential of utilizing oscillatory flow to selectively separate cell based on their mechanical attributes, which could be used in biological and medical applications. 

Our findings in Figure \ref{fig:croissant_transition_streamlines} show that the flow patterns are directly influenced by the dynamics of the RBC. Under steady-state flow, the extracellular flow dynamics were observed to behave differently near the RBC for the croissant and slipper shapes. In particular, the flow around the steady croissant shape was found similar to that of a rigid sphere \cite{Lee2010}, in which the flow streamlines move nearly symmetrically inwards and outward from the cell in the upstream and downstream sides, respectively. In contrast, for the slipper shape a fully-closed vortex ring more known as “bolus” was observed downstream the cell. Similar results were obtained using experimental Particle Tracking Velocimetry (PTV). Furthermore, our results in Figure \ref{fig:RBC_oscillatory_streamlines} suggest that it is possible to control the extracellular flow pattern by adjusting the inflow waveform. The extracellular flow has been found to play an important role in drug delivery strategies\cite{Yaya2021} due to its potential use of particle trapping. Therefore, our results suggest that controlling the inflow waveform either by adjusting the peak flow $\psi_f$ or the shape of the waveform ($T_f$) might lead to the desired effects in delivering small particles (e.x therapeutic nano-particles) to the cells. 
\section{Conclusion}
Transient dynamics of Red Blood Cells (RBC) in confined channels under oscillatory flows are investigated using our continuum-particle approach \cite{akerkouch2021}. Our results revealed that the dynamics of RBCs are complex with different shape modes that are beyond the usually observed croissant and slipper modes. Our results indicate that the extracellular flow pattern around the RBC is dependent on the RBC shape. Our results suggest that the oscillatory flow can be used to control and manipulate the dynamics of RBC by adapting appropriate flow waveform. Our specific conclusions are:

\begin{itemize}
     \item The RBC can transform into a variety of shapes such as multilobes, trilobes and hexalobes by varying the sinusoidal waveform even when it is subjected to a relatively weak flow shear rate ($\overline{\dot{\gamma_f}} \le 200~s^{-1}$) and sufficient channel confinement $\chi = 0.65$.

    \item Simple shapes such as croissant, slipper, and rolling discocyte appear when the RBC is subjected to all waveforms. However, complex shapes such as rolling stomatocyte, trilobes, and hexalobes appeared only under specific conditions. The appearance of a specific shape depends on the inlet waveform $(I)$. In our study, the RBC transitions into 8 shapes under the reversible waveform $(I_1)$, and into 5 shapes under the irreversible waveform $(I_2)$. Therefore, it is possible to attain a certain shape using an appropriate waveform.
    
    \item Under the reversible flow waveform, the axial displacement of the RBC is rather minimal. However, the lateral displacements are significantly large. Under the irreversible flow waveform, the RBC experiences a large axial displacement but small lateral displacements.
    
    \item The maximum lateral displacement of the RBC during its propagation depends on the initial radial shift ($r$). This maximum value is also dependent on the asymmetry of the flow waveform ($I$).     
    
    \item The extracellular flow surrounding the RBC depends on its morphological shape. The flow pattern is thus distinct and unique for each shape. 
\end{itemize}

\begin{acknowledgments}
This work is supported by the NSF grant number 1946202 ND-ACES and a start-up package of Trung Le from North Dakota State University. The authors acknowledges the use of computational resources at the Center for Computationally Assisted Science and Technology CCAST-NDSU, which is supported by the NSF MRI 2019077. The authors also received allocation CTS200012 from the Extreme Science and Engineering Discovery Environment (XSEDE). We acknowledge the financial support of  NIH-2P20GM103442-19A1 to train undergraduate students in Biomedical Engineering.
\end{acknowledgments}

\section*{Data Availability Statement}
The data that support the findings of this study are available from the corresponding author upon reasonable request.

\bibliography{aipsamp}

\clearpage

\clearpage
\begin{table}
\centering
\begin{tabular}[t]{lccc}
\toprule
Parameters~~~~ & ~~~~DPD value~~~~ & ~~~~Physical unit~~~~ & ~~~~Physical value\\
\hline
Bead  & 1  & $N_m$ & 3 $H_2 O$\\
$r_c$  & 1  & $(\rho N_m V)^\frac{1}{3}$ & $6.45~\AA$\\
$m$  & 1  & $\frac{N_m M}{N_A}$ & $8.98\times 10^{-23}~g$\\
$\rho$  & 3  & $\frac{\rho N_m M}{N_A r_c^3}$ & $996.3~kg~m^{-3}$\\
$\delta t$  & 0.01  & $\tau : \delta t r_c\sqrt{\frac{m}{k_B T}}$ & $1~ps$\\
$\mu_{cytosol} (\gamma = 116.4)$  & $4.11 \pm 0.1$  & $\frac{\eta \tau k_B T}{r_c^3}$ & $0.006~Pa~s$\\
\bottomrule
\end{tabular}
\caption{Relationship between DPD parameters and the physical units for viscosity ratio $\lambda = 5$. $N_m$, $m$, $\delta t$ and $\mu_{cytosol}$ correspond to the number of molecules in one bead, mass, time step and dynamic viscosity of the cytosol, respectively. $V$ is the volume of the water molecule $(30~\AA)$, $M$ is the molar weight of water $(18~g~mol^{-1})$ and $N_A = 6.0221415 \times 10^{23}$ is the Avogardo’s constant. The definitions of the parameters $k_B T$, $\gamma$, $\rho$ and $r_c$ are explained in Table \ref{tab:RBC_parameters} and section \ref{sec:dpd_to_physical}.}
\label{tab:dpd_fluid_visc}
\end{table}%

\clearpage
\begin{table}[H]
\centering
\begin{tabular}{ll}
    \toprule
    \multicolumn{2}{c}{RBC physical parameters}\\
    \hline    
    RBC diameter $(D_0)$   & ~~~~~~$7.82~\mu m$\\
    RBC area $(A_{0}^{tot})$   & ~~~~~~$135.0 \times 10^{-12}~m^2$\\
    RBC volume $(V_{0}^{tot})$ & ~~~~~~$94.0 \times 10^{-18}~m^3$\\
    Elastic shear modulus $(\mu_0)$ & ~~~~~~$6.3 ~\mu N/m$\\
    Young's modulus $(Y)$ & ~~~~~~$18.9 ~\mu N/m$ \\
    Bending rigidity $(k_b)$ & ~~~~~~$3.0 \times 10^{-19}~J$ \\
    Membrane viscosity $(\eta_m)$ & ~~~~~~$22.0 \times 10^{-3}~Pa~s$ \\
    Boltzmann's constant $(k_B)$ & ~~~~~~$1.380649 \times 10^{-23}~m^2~kg~s^{-2}~K^{-1}$\\
    Temperature  $(T)$ & ~~~~~~$298~K$\\
    \bottomrule
\end{tabular}
\caption{The physical parameters describing the RBC characteristics}
\label{tab:RBC_parameters}
\end{table}

\clearpage
\begin{table}[H]
\centering
\begin{tabular}{ccccccc}  
    \toprule
        $N_v$~~ & ~~$D_{0}^M$~~ & ~~$l_{0}$~$(m)$~~ & ~~$l_{max}$~$(m)$~~ &  ~~$p$~$(m)$~~ & ~~$k_p$~$(N~m^2)$~~ & ~~ $\theta_{0}$~$(deg)$\\
    \midrule
         $500$~~  & ~~$8.07$~~  &  ~~$5.5614 \times 10^{-7}$~~   &  ~~$1.2235 \times 10^{-6}$~~ & ~~ $1.9933 \times 10^{-9}$~~ & ~~ $1.2626\times 10^{-24}$ ~~& ~~$6.86$ \\
        $1000$~~  & ~~$8.07$~~  &  ~~$3.7992 \times 10^{-7}$~~   &  ~~$8.3582 \times 10^{-7}$~~ & ~~ $2.9179  \times 10^{-9}$~~ & ~~ $4.0252\times 10^{-25}$ ~~& ~~$4.69$ \\
       $3000$~~  & ~~$8.07$~~  &  ~~$2.2818 \times 10^{-7}$~~   &  ~~$5.0199 \times 10^{-7}$~~ & ~~ $4.8584 \times 10^{-9}$~~ & ~~ $8.7205\times 10^{-26}$ ~~& ~~$2.82$ \\
       $9000$~~  & ~~$8.12$~~  &  ~~$1.3035 \times 10^{-7}$~~   &  ~~$2.8678 \times 10^{-7}$~~ & ~~ $8.5044 \times 10^{-9}$~~ & ~~ $1.6259\times 10^{-26}$ ~~& ~~$1.61$ \\
       $24,472$~~  & ~~$8.26$~~  &  ~~$7.5331 \times 10^{-8}$~~   &  ~~$1.6573 \times 10^{-7}$~~ & ~~ $1.4716 \times 10^{-8}$~~ & ~~ $1.6589\times 10^{-27}$ ~~& ~~$0.93$ \\
    \bottomrule
    \end{tabular}
    \caption{Coarse-grained parameters for the RBC membrane model for different numbers of vertices $N_v$. The definitions of the parameters $D_0^M$, $l_0$, $l_{max}$, $p$, $k_p$ and $\theta_0$  are explained in sections \ref{sec:rbc_membrane_model} and \ref{sec:Scaling}. The corresponding values of Young's modulus, global area, local area and volume constraints in DPD units are $Y^M = 392.5$, $k_a = 4900$, $k_d = 100$, $k_v = 5000$, respectively. Other parameters are $\alpha=1$ and $\eta_m^M = 1.8$.}
     \label{tab:RBC_DPD_Parameters}
     \end{table}


\clearpage
\begin{table}[H]
\centering
\begin{tabular}{lcccc}
\toprule
Channel~~~~ & ~~~~$L_x \times L_y \times L_z~(\mu m)$ & $N_i \times N_j \times N_k$ & $\Delta x \times \Delta y \times \Delta z~(\mu m)$ & ~~~~$\chi$\\
\hline
1~~~~       &  ~~~~$90 \times 12 \times 10$~~~~ & ~~~~$151 \times 101 \times 101$~~~~  &  ~~~~$ 0.6 \times 0.12 \times 0.1$ ~~~~& ~~~~0.65\\ 

2~~~~       & ~~~~$80 \times 16 \times 16 $~~~~ & ~~~~$151 \times 101 \times 101$~~~~  &  ~~~~$ 0.54 \times 0.16 \times 0.16$~~~~ & ~~~~0.4\\

3~~~~       & ~~~~$80 \times 21 \times 21$~~~~ & ~~~~$151 \times 151 \times 151$~~~~  &  ~~~~$ 0.54 \times 0.14 \times 0.14$~~~~ & ~~~~0.3\\
\bottomrule
\end{tabular}
\caption{Different channel geometries and their associated computational grids to simulate the dynamics of RBC in fluid flows. The channels have rectangular cross-sections of size $L_x$, $L_y$, $L_z$ along the axial, spanwise, and vertical directions, respectively. $N_i$, $N_j$ and $N_k $ are respectively the number of grid points in $x$, $y$, $z$ directions. $\chi$ is the channel confinement, which is defined in section \ref{sec:FSI_configs}.}
\label{tab:Channels}
\end{table}%

\clearpage
\begin{table}
    \centering
    \begin{tabular}{cccccccc}
    \hline
        Case~~~~  &  ~~~~Inflow~~~~  & ~~~~$\psi_f~(mm/s)$ & ~~~~$\psi_b~(mm/s)$~~~~ & ~~~~$\bar{\dot{\gamma}}_f (s^{-1})$~~~~ &  ~~~~$r (\mu m)$ & $Re_f$   & $Ca^f$\\
        \hline
        $I_0U_3r_1\chi_1$ &   $I_0$ &   2   & - &   200         & 0     & ~~~~$1.34 \times 10^{-2}$      & ~~~~$0.12$    \\
        $I_0U_4r_3\chi_1$ &   $I_0$ &   6   & - &   600         & 0.7   & ~~~~$4 \times 10^{-2}$         & ~~~~$0.37$ \\
     $I_s\psi_1 r_1\chi_2$ &   $I_s$ &  $1.05$  & $-0.27$ &   $66$         & 0  &  $7 \times 10^{-3}$  & ~~~~$0.1$ \\
     $I_s\psi_2 r_1\chi_2$ &   $I_s$ &  $1.56$  & $-0.39$ &  $98$         & 0  &  $1.04 \times 10^{-2}$  & ~~~~$0.15$\\
     $I_s\psi_3 r_1\chi_2$&   $I_s$ &  $2.1$  &  $-0.53$ & $132$          & 0  &  $1.4 \times 10^{-2}$  & ~~~~$0.2$ \\
     $I_s\psi_4 r_1\chi_3$ &   $I_s$ &  $2.17$  & $-0.55$ &  $104$          & 0  &  $1.45 \times 10^{-2}$ &~~~~ $0.1$ \\
     $I_s\psi_5 r_1\chi_3$&   $I_s$ &  $3.25$  & $-0.82$ &  $155$          & 0  &  $2.17 \times 10^{-2}$ & ~~~~$0.15$ \\
     $I_s\psi_6 r_1\chi_3$ &   $I_s$ &  $4.34$  & $-1.09$ &  $207$          & 0  &  $2.9 \times 10^{-2}$ & ~~~~$0.2$ \\
        \hline
    \end{tabular}
    \caption{Summary of the validation cases  under constant shear rates ($I_0U_3r_1\chi_1$ and $I_0U_4r_3\chi_1$), and stepwise oscillatory flows ($I_s\psi_1r_1\chi_2$, $I_s\psi_2r_1\chi_2$, $I_s\psi_3r_1\chi_2$ and $I_s\psi_4r_1\chi_3$, $I_s\psi_5r_1\chi_3$, $I_s\psi_6r_1\chi_3$). The stepwise oscillatory flows with the forward $(\psi_f)$, backward $(\psi_b)$ velocities and the forward Capillary number are defined in section \ref{sec:oscillatory_shear_rate}. The maximum shear rate $(\bar{\dot{\gamma}}_f)$ and the maximum Reynolds number $(Re_f)$ are defined in section \ref{sec:constant_shear_rate}. The definition of the RBC's radial shift $r$ is shown in Figure \ref{fig:simulation_setup}b.}
    \label{tab:validation_summary}
\end{table}

\clearpage
\begin{table}[ht]
\centering
\caption{The controlling parameters of the pulsatile waveforms ($I_1$, $I_2$, $I_3$ and $I_4$). The waveforms are characterized by the intervals of the forward $(T_f)$, rest $(T_r)$ and backward $(T_b)$ periods. The shapes of the waveforms are shown in Figure \ref{fig:waveform_types}.}
\begin{tabular}[t]{l|cccc}
\hline
\multicolumn{1}{c|}{\diagbox[width=\dimexpr \textwidth/4+2\tabcolsep\relax, height=1.2cm]{Waveforms}{Time}} & ~~~~$T_f~(ms)$~~~~ & ~~~~$T_r~(ms)$~~~~ & ~~~~$T_b~(ms)$\\
\hline
$I_1$ & $20$ & $10$ & $20$\\
$I_2$ & $25$ & $12.5$ & $12.5$\\
$I_3$ & $27.3$ & $9$ & $13.7$ \\
$I_4$ & $28.6$ & $7.1$ & $14.3$\\
\hline
\end{tabular}
\label{tab:physical_time}
\end{table}%

\clearpage
\begin{table}
    \centering
    \begin{tabular}{cccc}
    \hline
    Subscript~~~     &  ~~~Inflow waveform~~~ & ~~~$\psi_f (mm/s)$~~~ & ~~~ Radial shift $r (\mu m)$\\
    \hline
        1         &   $I_1$ & $U_1 = 1$ & $r_1 = 0$ \\
        2         &   $I_2$ & $U_2 = 1.5$ & $r_2 = 0.4$ \\
        3         &   $I_3$ & $U_3 = 2$ & $r_3 = 0.7$ \\
        4         &   $I_4$ & $U_4 = 6$ & --  \\
    \hline
     \end{tabular}
    \caption{The summary of the combinatoric configurations for steady and pulsatile flow simulations. The combination of the waveform type, the forward flow velocity, the radial shift, and the channel confinement results in the simulation configurations of $I_m U_n r_p \chi_x$. Here $m=1,~2,~3,~and~4$, $n=1,~2,~3~and~4$, $p = 1,~2~and~3$ and $x = 1,~2~and~3$. The profile of the inflow waveforms ($I_1$, $I_2$, $I_3$ and $I_4$) are shown in Figure \ref{fig:waveform_types}. The peak forward flow velocity $U$ ( Figure \ref{fig:waveform_types}a) varies from 1 to 6 mm/s. The radial shift of the RBC centroid along the bisector of the $y-z$ plane at the initial time is defined in Figure \ref{fig:simulation_setup}b.}
    \label{tab:summary_factors}
\end{table}

\clearpage
\begin{table}[h]
    \centering
    \includegraphics[width=1.1\linewidth]{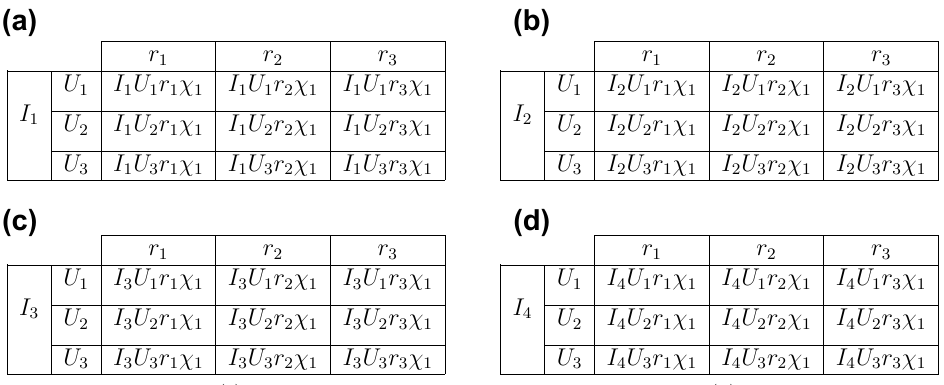}
    \caption{Summary of the 36 sinusoidal flow cases in section \ref{sec:sinusoidal_flow_simulations} with a confinement of $\chi_1 = 0.65$. Tables $(a)$, $(b)$, $(c)$, or $(d)$ each consists of 9 possible combinations between the peak forward flow $U$ and the radial shift $r$ for each type of waveform $I_1$, $I_2$, $I_3$ and $I_4$, respectively. The exact numeric value of $U_1, U_2, U_3$ and $r_1, r_2, r_3$ are shown in Table \ref{tab:summary_factors}.}
    \label{tab:all_our_oscillaotry_cases}
\end{table}

\clearpage
\begin{table}[h]
\begin{tabular}{ccccccccccc}
\toprule
\multirow{2}{*}{Waveform $(I_1)$} & \multicolumn{1}{c}{C} & \multicolumn{2}{c}{S} & \multicolumn{1}{c}{CM} & \multicolumn{1}{c}{M} & \multicolumn{1}{c}{T} & \multicolumn{1}{c}{RS} & \multicolumn{1}{c}{EM} & \multicolumn{2}{c}{RD}\\
\cmidrule(rl){2-2} \cmidrule(rl){3-4} \cmidrule(rl){5-5} \cmidrule(rl){6-6} \cmidrule(rl){7-7} \cmidrule(rl){8-8} \cmidrule(rl){9-9} \cmidrule(rl){10-11}
 & ~~~~{$r_1$}~~~~  & ~~~~{$r_2$}~~~~ & ~~~~{$r_3$}~~~~  & ~~~~{$r_1$}~~~~ & ~~~~{$r_1$}~~~~ & ~~~~{$r_1$}~~~~  & ~~~~{$r_1$}~~~~ & ~~~~{$r_1$}~~~~ &~~~~{$r_2$}~~~~ & ~~~~{$r_3$}~~~~\\
\midrule
$U_1$ & 0.32 & 0.3 & 0.3 & - & 0.75  & -  & 0.9  & -  & 1.25 & 1.25 \\
\noalign{\vskip 2mm}
$U_2$ & 0.21 & 0.2 & 0.2 & 0.5  & - & 0.8  & - & 1.2 & 1.21 &1.21 \\
\noalign{\vskip 2mm}
$U_3$ & 0.21 & 0.2 & 0.2 & 0.5 &  - & 0.8 & - & 1.2 & 1.21 & 1.21 \\
\bottomrule
\end{tabular}
\caption{Summary of the RBC morphology transition sequences recorded at different time instances $\frac{t}{T}$ under $I_1$ waveform, and different flow velocities $U_1, U_2, U_3$ and radial shift $r_1, r_2, r_3$. Here, the time instances represent the first time the RBC deformed shape appeared. The acronyms $C$, $S$, $CM$, $M$, $T$, $RS$, $EM$ and $RD$ represent the croissant, slipper, complex multilobes, multilobes, trilobes, rolling stomatocye, elongated multilobes and rolling discocyte, respectively. The exact numeric value of $U_1, U_2, U_3$ and $r_1, r_2, r_3$ are shown in Table \ref{tab:summary_factors}.}
\label{tab:I1}
\end{table}

\clearpage
\medskip
\begin{table}[h]
\begin{tabular}{ccccccccc}
\toprule
\multirow{2}{*}{Waveform $(I_2)$} & \multicolumn{1}{c}{C} & \multicolumn{2}{c}{S} & \multicolumn{1}{c}{CM} & \multicolumn{2}{c}{EM} & \multicolumn{2}{c}{RD} \\
\cmidrule(rl){2-2} \cmidrule(rl){3-4} \cmidrule(rl){5-5} \cmidrule(rl){6-7} \cmidrule(rl){8-9}
 & ~~~~{$r_1$}~~~~ & ~~~~{$r_2$}~~~~ & ~~~~{$r_3$}~~~~ & ~~~~{$r_1$}~~~~ & ~~~~{$r_2$}~~~~ & ~~~~{$r_3$}~~~~  & ~~~~{$r_2$}~~~~ & ~~~~{$r_3$}~~~~\\
\midrule
$U_1$ & 0.27 & 0.29 & 0.29 & 0.9 & - & - & 1.33 & 1.33\\
\noalign{\vskip 2mm}
$U_2$ & 0.2 & 0.22 & 0.22 & 1.06 & - & - & 1.3 & 1.3\\
\noalign{\vskip 2mm}
$U_3$ & 0.2 & 0.22 & 0.22 & 1.06  & 1.16 & 1.16  & 1.3 & 1.3 
\\
\bottomrule
\end{tabular}
\caption{Summary of the RBC morphology transition sequences recorded at different time instances $\frac{t}{T}$ under $I_2$ waveform, and the different flow velocities $U_1, U_2, U_3$ and initial placements $r_1, r_2, r_3$. Here, the time instances represent the first time the RBC deformed shape appeared. The acronyms $C$, $S$, $CM$, $EM$ and $RD$ represent the croissant, slipper, complex multilobes, elongated multilobes and rolling discocyte, respectively. The exact numeric value of $U_1, U_2, U_3$ and $r_1, r_2, r_3$ are shown in Table \ref{tab:summary_factors}.}
\label{tab:I2}
\end{table}

\clearpage
\medskip
\begin{table}[h]
\begin{tabular}{cccccccccc}
\toprule
\multirow{2}{*}{Waveform $(I_3)$} & \multicolumn{1}{c}{C} & \multicolumn{2}{c}{S} & \multicolumn{1}{c}{CM} & \multicolumn{1}{c}{T} & \multicolumn{2}{c}{EM} & \multicolumn{2}{c}{RD}\\
\cmidrule(rl){2-2} \cmidrule(rl){3-4} \cmidrule(rl){5-5} \cmidrule(rl){6-6} \cmidrule(rl){7-8} \cmidrule(rl){9-10}
 & ~~~~{$r_1$}~~~~ & ~~~~{$r_2$}~~~~ & ~~~~{$r_3$}~~~~ & ~~~~{$r_1$}~~~~ & ~~~~{$r_1$}~~~~ & ~~~~{$r_2$}~~~~ & ~~~~{$r_3$}~~~~& ~~~~{$r_2$}~~~~ & ~~~~{$r_3$}~~~~ \\
\midrule
$U_1$ & 0.34 & 0.27 & 0.27  & 0.58  & -  & - & - & 1.35 & 1.35 \\
\noalign{\vskip 2mm}
$U_2$ & 0.28  & 0.22 & 0.22 & 0.62 & 0.94 & - & - & 1.32 & 1.32 \\
\noalign{\vskip 2mm}
$U_3$ & 0.28 & 0.22 & 0.22 & 0.62 & - & 1.18 & 1.18 & 1.32 & 1.32 \\
\bottomrule
\end{tabular}
\caption{Summary of the RBC morphology transition sequences recorded at different time instances $\frac{t}{T}$ under $I_3$ waveform, and the different flow velocities $U_1, U_2, U_3$ and initial placements $r_1, r_2, r_3$. Here, the time instances represent the first time the RBC deformed shape appeared. The acronyms $C$, $S$, $CM$, $T$, $EM$ and $RD$ represent the croissant, slipper, complex multilobes, trilobes, elongated multilobes and rolling discocyte, respectively. The exact numeric value of $U_1, U_2, U_3$ and $r_1, r_2, r_3$ are shown in Table \ref{tab:summary_factors}.}
\label{tab:I3}
\end{table}

\clearpage
\medskip
\begin{table}[h]
\begin{tabular}{cccccccccc}
\toprule
\multirow{2}{*}{Waveform $(I_4)$} & \multicolumn{1}{c}{C} & \multicolumn{2}{c}{S} & \multicolumn{1}{c}{CM} & \multicolumn{2}{c}{EM} & \multicolumn{2}{c}{RD} & \multicolumn{1}{c}{HX} \\
\cmidrule(rl){2-2} \cmidrule(rl){3-4} \cmidrule(rl){5-5} \cmidrule(rl){6-7} \cmidrule(rl){8-9} \cmidrule(rl){10-10}
 & ~~~~{$r_1$}~~~~ & ~~~~{$r_2$}~~~~ & ~~~~{$r_3$}~~~~ & ~~~~{$r_1$}~~~~ & ~~~~{$r_2$}~~~~ & ~~~~{$r_3$}~~~~ & ~~~~{$r_2$}~~~~ & ~~~~{$r_3$}~~~~ & ~~~~{$r_1$}~~~~\\
\midrule
$U_1$ & 0.32 & 0.31 & 0.31 & 0.6 & - & - & 1.27 & 1.27 & -\\
\noalign{\vskip 2mm}
$U_2$ & 0.27 & 0.24 & 0.24 & 0.56 & - & - & 1.24 & 1.24 & 1.15\\
\noalign{\vskip 2mm}
$U_3$ & 0.27 & 0.24 & 0.24 & 0.56& 1.15 & 1.15 & 1.24 & 1.24 & -\\
\bottomrule
\end{tabular}
\caption{Summary of the RBC morphology transition sequences recorded at different time instances $\frac{t}{T}$ under $I_4$ waveform, and the different flow velocities $U_1, U_2, U_3$ and initial placements $r_1, r_2, r_3$. Here, the time instances represent the first time the RBC deformed shape appeared. The acronyms $C$, $S$, $CM$, $EM$, $RD$ and $HX$ represent the croissant, slipper, complex multilobes, elongated multilobes, rolling discocyte and hexalobes, respectively. The exact numeric value of $U_1, U_2, U_3$ and $r_1, r_2, r_3$ are shown in Table \ref{tab:summary_factors}.}
\label{tab:I4}
\end{table}


\clearpage

\clearpage


\clearpage
 
\end{document}